\documentclass[12pt]{article}

\usepackage{axodraw} %%%%%%%%%%%%%%%%%questo va messo nella directory
\usepackage{graphics}
\usepackage{amssymb}
\def\eq#1{Eq.~(\ref{#1})}
\newcommand{\secn}[1]{Section~\ref{#1}}

\newcommand{\fig}{Fig.~\ref}
\newcommand{\nl}{\nonumber \\}

\def\beq{\begin{equation}}
\def\eeq{\end{equation}}
\def\beqa{\begin{eqnarray}}
\def\eeqa{\end{eqnarray}}
\newcommand{\Tr}{{\rm Tr}}
\newcommand{\sect}[1]{\setcounter{equation}{0}\section{#1}}

\renewcommand{\a}{\alpha}

\renewcommand{\l}{\lambda}
\newcommand{\e}{\epsilon}

\renewcommand{\thefootnote}{\fnsymbol{footnote}}
\def\bet{\begin{tabular}}
\def\eet{\end{tabular}}

\begin{document}
\begin{titlepage}
\rightline{DFTT 26/00}
\rightline{LPTENS-00-41}
\rightline{\hfill December 2000}

\vskip 2.5cm

\centerline{\Large \bf Systematics of one--loop Yang--Mills diagrams}
\centerline{\Large \bf from bosonic string amplitudes} 
 
\vskip 2cm

\centerline{\bf Alberto Frizzo and Lorenzo Magnea\footnote{e-mail: 
magnea@to.infn.it}}
\centerline{\sl Dipartimento di Fisica Teorica, Universit\`a di Torino}
\centerline{\sl and I.N.F.N., Sezione di Torino}
\centerline{\sl Via P.~Giuria 1, I--10125 Torino, Italy}
 
\vskip .5cm
 
\centerline{\bf Rodolfo Russo}
\centerline{\sl Laboratoire de Physique Th\'eorique de
 l'Ecole Normale Sup\'erieure }
\centerline{\sl 24 rue Lhomond, {}F-75231 Paris Cedex 05, France}
 
\vskip 2cm
 
\begin{abstract}

We present a general algorithm to compute off--shell, one--loop
multigluon Green functions using bosonic string amplitudes. We
identify and parametrize the regions in the space of moduli of
one--loop Riemann surfaces that contribute to the field theory limit
of string amplitudes.  Each of these regions can be precisely
associated with a Feynman--like scalar graph with cubic and quartic
vertices, whose lines represent the joint propagation of ghosts and
gluons.  We give a procedure to compute the contribution of each graph
to a gluon Green function, for arbitrarily polarized off--shell
gluons, reducible and irreducible diagrams, planar and non--planar
topologies. Explicit examples are given for up to six gluons.

\end{abstract}

\end{titlepage}

\newpage
\renewcommand{\thefootnote}{\arabic{footnote}}
\setcounter{footnote}{0}
\setcounter{page}{1}

\sect{Introduction}
\label{intro}

The fact that string techniques can be used to compute on--shell
scattering amplitudes in Yang--Mills theory at tree
level~\cite{nesche72,manpa} and one loop \cite{beko,bebo,bernrev} has been
known for many years, and has inspired the developments of novel
techniques also in field theory~\cite{schu}.  There are several
reasons, both from the point of view of strings and in view of further
applications to Yang--Mills theory, that lead to consider the
extension of this formalism to off--shell correlation functions.

In the context of string theory, it is generally accepted that the
framework developed so far for the calculation of scattering
amplitudes corresponds to a first--quantized theory, and is properly
suited only for on--shell quantities; a consistent extension off the
mass--shell of string--derived amplitudes in the field theory limit
would provide a non--trivial testing ground for proposals to compute
off--shell amplitudes in the full string theory~\cite{nap}. Recently,
similar techniques have been applied also to a non--commutative
generalization of field theory~\cite{noncom}.

In the context of particle theory, there are several points of
interest in going off--shell. First, only with off--shell gluons one
can make a definite identification of the gauge selected by string
theory; this issue was addressed shortly after the first complete
one--loop computations~\cite{bedu}, and a combination of gauges that
would reproduce the simplifying features of the string calculation was
identified. This identification was confirmed in
Refs.~\cite{letter,big}, where a consistent prescription to continue
off--shell the string amplitude was proposed, and it was shown that
divergent parts of one--loop diagrams were consistent with the
identification made in~\cite{bedu}. A general argument detailing how
the string amplitude splits into particle graphs, and proving that all
such off--shell graphs correspond to a definite gauge, including
finite contribution, was however still missing. A second motivation is
provided by the desire to extend the string formalism to diagrams with
more loops: although partial progress has been made, in the context of
scalar amplitudes \cite{2loop,scal,mape}, and for the simplest gluon
diagrams \cite{maru}, tackling a general two--loop gluon amplitude
has proved difficult \cite{schmk}. The main source of this difficulty
is the fact that there are several simplifying features of one--loop
string calculations that are not preserved by a two--loop
generalization: at one loop, for example, it is possible to use
partial integration at the string level to eliminate completely the
regions of integration corresponding to four point vertices, and
furthermore the topology of the graphs contributing non--trivially to
the amplitude is essentially unique. At two loops, it becomes
necessary to understand in detail how string moduli space degenerates
into a set of graphs in the field theory limit, and the study of
scalar diagrams~\cite{scal} has shown that the reconstruction of
Feynman graphs from contributions arising from different regions of
moduli space is non--trivial. To solve the problems that arise at two
loops, it is then necessary to develop a more systematic approach to
the study of the field theory limit; in particular, it is necessary to
identify precisely the regions of integration corresponding to
particle graphs of various topologies and vertex structures. This can
be done consistently only off the mass shell, with a precise
identification of the gauge choice, because in Yang--Mills theory the
contributions of different graphs mix when the gauge is changed, and
the differences only cancel when one sums graphs to reconstruct a
gauge--invariant quantity. As a final motivation, it should be
emphasized that in Yang--Mills theory and in QCD there are many
quantities of theoretical and phenomenological interest that are
defined off the mass shell; it would be interesting to apply string
techniques to these quantities as well.

In the present paper we will perform a systematic study, of the kind
just described, for one--loop Yang--Mills amplitudes. Starting from
the string master formula for one--loop gluon amplitudes, we will give
a precise characterization of all the regions of integration
contributing to the field theory limit. Each such region corresponds
to a Feynman--like graph, where, however, the color algebra, the
Lorentz algebra, and the loop momentum integration have already been
performed; because of this, we will refer to such a graph as a `scalar
graph'. One--loop scalar graphs correspond to sums of Feynman
diagrams, since both gluons and ghosts circulate in the loop. Scalar
graphs in general contain both cubic and quartic vertices, and can be
one--particle reducible or irreducible, as well as planar or
non--planar. We will give examples of each kind: in particular, we
will consider in detail the case of diagrams with quartic vertices,
which arise from integration over finite regions in string moduli
space, and in some cases require a regularization of infrared
singularities arising from the propagation of bosonic string tachyons
in intermediate channels.

The string master formula, when applied to off--mass--shell,
non--transverse external gluons, depends on a set of projective
transformations (one for each external state), defining a local
coordinate system on the world--sheet around the point of insertion of
each external particle. At one loop, geometrical arguments lead to an
essentially unique choice of these projective transformation that
preserves the global properties of the geometric objects appearing in
the string master formula\footnote{A choice of this kind was proposed
in Ref.~\cite{nap}, where results similar to ours were obtained for
the two-- and the three--point functions.}. We will show that, in each
of the cases examined, this choice leads to a well--defined off--shell
continuation of the amplitude, which corresponds to the combination of
gauges described in Refs.~\cite{bedu} and~\cite{big}. The
correspondence is apparent at the level of the integrands when the
amplitude is expressed in terms of Schwinger or Feynman parameters, so
it can be checked case by case without having to perform any
integration. Since the results are valid for external gluons with
arbitrary mass and polarization, polarization vectors could be dropped
and the results could be read directly as the values of the
corresponding truncated Green functions, expressed as functions of the
external momenta.

We emphasize that, although a precise correspondence between scalar
graphs arising from string theory and sets of diagrams in field theory
is established, it remains true that the string method is vastly
convenient from a practical point of view, essentially because it
generates automatically the answer for the sum of a subset of diagrams
{\it after} loop momentum integration, directly in terms of external
momenta and polarizations, bypassing all intermediate stages of the
calculation, in which typically hundreds of thousands of terms are
generated. We emphasize also that, in its present form, the method
lends itself to be completely automated, in the sense that all
expressions for the integrands of all relevant scalar graphs can be
machine--generated in a short time starting from the basic string
ingredients. The only obstacle remaining on the way to a complete
calculation is the computation of scalar integrals, something for
which this approach can offer no help.

The paper is organized as follows.  In \secn{General}, we introduce
the necessary ingredients of the string formalism, we explain the role
of projective transformations, and give the geometric prescription for
the off--shell continuation. In \secn{GeneralFT} we describe the
general features of the procedure used to take the field theory limit,
introducing the mapping between string moduli and Schwinger parameters
in field theory. In \secn{Reg1}, we focus on the regions in moduli
space corresponding to diagrams with cubic vertices only; we give
examples of 3 and 4 point diagrams and specifically of a non--planar
contribution.  In \secn{Reg2}, we show how to treat quartic vertices,
defining the necessary regularization prescriptions. In \secn{Reg3} we
turn our attention to reducible diagrams, and verify the non--trivial
gauge choice performed by the string in this case.  In \secn{syst} we
summarize our calculational method, which is now as general and
systematic as conventional Feynman rules, at least for one--loop
correlation functions, and we give a concrete example of the
application of the method to a more complicated graph, computing a
contribution to the six--gluon one--loop amplitude.  Finally, in
\secn{concl}, we draw our conclusions and discuss possible
developments and applications.

\sect{The one--loop off--shell master formula}
\label{General}

Open bosonic string theory is the simplest string model containing a
massless vector state. Of course, the whole spectrum consists of an
infinite tower of states with masses proportional to the string
tension $T$; however, all the contributions to scattering amplitudes
coming from massive modes disappear in the low--energy limit, $T\to
\infty$, and the dynamics of the surviving light modes can be
effectively described by a field theory. As discussed in
Ref.~\cite{scal}, it is possible to reproduce different field theories
in the infinite tension limit by choosing different matching
conditions for the coupling. Here we will be concerned with the
Yang--Mills field theory limit (first discussed in Ref.~\cite{bebo}), 
which is obtained by allowing the circulation of only massless states 
in the loop.

There are two potential technical problems in the computation of the
Yang--Mills field theory limit: the fact that the ground state of the
spectrum of the bosonic string is a tachyon, and the fact that string
amplitudes are computed in the critical dimension $d = d_c =
26$. Neither problem affects the result: it turns out that tachyonic
contributions to gluon amplitudes correspond to infrared divergences
that can be easily identified and regulated by hand, and furthermore
one can continue the string amplitude to arbitrary values of $d \neq
26$, obtaining in fact the correct result for the dimensionally
regularized field theory of interest.

The continuation to arbitrary $d$ can be justified form the point of
view of string theory by constructing a simple, though consistent,
string model, having as low--energy limit the pure $d$--dimensional
Yang--Mills theory~\cite{Gomis:2000bn}. It is sufficient to break the
full $26$--dimensional Lorentz invariance by imposing different
boundary conditions on the open string along different
directions. Since we want open strings to be free to move in $d$ of
the original $26$ dimensions, we split our space time as ${\cal
M}_{26} = M_d \times {\cal M}_{26 - d}$, where $M_d$ is
$d$--dimensional Minkowsky space; then we choose Neumann boundary
conditions along $M_d$ and Dirichlet boundary conditions along ${\cal
M}_{26 - d}$; in other words, we consider a bosonic $D(d-1)$--brane.
It is well known that the massless sector of such open strings also
contains states which are scalars from the $d$--dimensional point of
view. In order to decouple these unwanted scalars, one can choose
${\cal M}_{26 - d}$ to be the orbifold ${\bf R}^{26 - d}/Z_2$, where
the $Z_2$ orbifold operation is simply the reflection in ${\bf R}^{26
- d}$ around the origin.  The orbifolded theory must contain only the
open string states which are invariant under $Z_2$. By choosing a
trivial action of this $Z_2$ operation on the Chan--Paton factor, it
is easy to see that all the massless scalar are projected out by the
orbifold operation, and one is left with a pure $d$--dimensional
Yang--Mills theory (in the $D$--brane language, we are using bosonic
`fractional branes'). The use of this kind of $D$--branes, present in
orbifolded theories, is a quite general technique for generating pure
Yang--Mills theory also in the supersymmetric case.

Of course, the string master formula one derives in the orbifold model
would be more complicated than the one presented here, in
\eq{hmastac}.  However, as long as one considers scattering amplitudes
only among vector states, the differences conspire precisely, in the
field theory limit, to change the effective value of the dimension
from $d_c = 26$ to $d$~\cite{Gomis:2000bn}.  Thus, computing the
string amplitudes from \eq{hmastac} directly in $d$ dimensions is, at
one loop and in the field theory limit, a shortcut perfectly
equivalent to the computation in the orbifolded model in the critical
dimension. The gluon amplitude computed this way automatically has the
correct $d$ dependence, and thus is given in dimensionally regularized
form. This also provides a simple criterion to identify tachyonic
contributions: they are in fact infrared divergent in the field theory
limit for all values of $d$, {\it i.e.}  they are not regularized by
dimensional continuation. Such contributions must be discarded, just
as the contributions of massive states, although they may leave finite
remainders after regularization, as we will see in \secn{Reg2}.

A further technical problem of special relevance to the present paper
is the fact that string amplitudes must be computed between states
satisfying the mass--shell condition, to preserve conformal invariance
on the string world--sheet. It turns out, however, that the effects of
off--shell continuation are parametrized in a simple way in the
context of the operator formalism \cite{copgroup}: the amplitude
acquires a dependence on a set of projective transformations defining
local coordinate systems around the locations of the insertions of
external gluons. As we will see, there is a simple, geometrically
motivated choice for these projective transformations, which
consistently yields the correct off--shell extension for Yang--Mills
amplitudes.

Since our aim here is to provide a self--contained method to calculate
gluon diagrams, we will not emphasize technical details of the string
formalism, which are available elsewhere \cite{copgroup,GSW}. We will
just give the general expression for the one--loop multigluon string
amplitude, and describe the geometric features necessary to define and
perform the field theory limit.

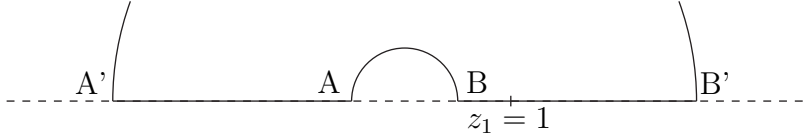
\begin{figure}
\begin{center}
\begin{picture}(300,150)(0,0)
\DashLine(0,10)(300,10){3}
\CArc(150,10)(20,0,180)
\CArc(150,10)(110,0,20)
\CArc(150,10)(110,160,180)
\Line(40,10)(130,10)
\Line(170,10)(260,10)
\Text(32,17)[]{A'}
\Text(122,17)[]{A}
\Text(178,17)[]{B}
\Text(268,17)[]{B'}
\Line(190,12)(190,8)
\Text(190,-2)[b]{$z_1=1$}
\end{picture}
\end{center}
\caption{The annulus in Schottky representation.}\label{scho}
\end{figure}

It should be kept in mind that string theory, although describing the
propagation of objects in a $d$--dimensional space time, has a
two--dimensional structure describing the world--sheet dynamics. Thus
the amplitudes for the scattering of string states, and the
ingredients entering in their expression, must have a meaning also
from the point of view of the two--dimensional theory. String
amplitudes must then be written in the language of Riemann surfaces,
in our case with boundaries and punctures representing the insertions
of external gluons. Using the operator formalism \cite{copgroup},
Riemann surfaces turn out to be represented in the Schottky
parametrization~\cite{copscho}, which is the one we will use. In
particular, one--loop open string amplitudes involve insertions of
gluon vertex operators on the boundaries of an annulus, which, in the
Schottky parametrization, is represented as in \fig{scho}.

%%%%%%%%%%%%%%%%%%%%%%%%%%%%%%%%%%%%%%%%%%%%%%%%%%%%%%%%%%%%%%%%%%%%%
%%%%%%%%%%%%%%%%%%%%%%%%%%%%%%%%%%%%%%%%%%%%%%%%%%%%%%%%%%%%%%%%%%

The boundaries of the annulus correspond to the segments $A A'$ and $B
B'$, whose endpoints must be identified. External gluons will be
inserted on the boundaries at locations $z_i$, often referred to as
`punctures'.  It is easy to see how the idea of adding loops is
implemented in this formalism. One starts from the upper half complex
plane (equivalent to the disk, representing the tree--level scattering
amplitude), and adds two circles with centers on the real axis, which
must then be identified via a projective transformation. Each loop is
thus characterized by three real parameters: the positions of the two
centers on the real axis and the common radius of the circles, which
fix respectively the positions and the width of the holes added to the
surface. The positions of the various circles are related to the fixed
points of the projective transformations under which the pairs of
circles are identified.  Usually the fixed points are denoted by $\xi$
and $\eta$, while the width of the holes is determined by the third
parameter characterizing the projective transformation, the multiplier
$k$. The fundamental building block entering the amplitudes is the
Green function of the two--dimensional Laplace operator defined on the
string world--sheet, denoted by ${\cal G}(z_i,z_j)$: it is the
correlator of two world--sheet bosons located at positions $z_i$ and
$z_j$ and of course depends on the parameters $(\xi,\eta,k)$ defining
the Riemann surface.

At this point we can write down the `master formula' for the one--loop
multi--gluon string amplitude, in terms of the Green function ${\cal
G}$.  If one does not impose the mass--shell and transversality
conditions on the external states, the Green function must be
evaluated in terms of the projective transformations associated with
the punctures, denoted by $V_i(z)$, which must satisfy
\beq
V_i(0) = z_i~, 
\label{local}
\eeq
in order to define local coordinates around $z_i$.  The master formula
is then given by
\beqa
& & \hspace{-1cm} A^{(1)} (1, \ldots, M) = {\rm Tr}(\lambda^{a_1} 
\ldots \lambda^{a_{m}}) \, 
{\rm Tr}(\lambda^{a_{m + 1}} \ldots \lambda^{a_M}) ~\frac{2^M ~g_S^{M}}{(2
\pi)^{d}} \, \left( 2 \a' \right)^{(M d - 2 M - 2 d)/4} \nl 
& \times &
\int \frac{\prod_{i = 1}^M d z_i}{d V_{abc}~ V'_i(0)}~
\frac{d k~d \xi~d \eta~}{k^2 (\xi - \eta)^2} 
\left(- \frac{\ln k}{2\pi} \right)^{- d/2}
\prod_{n = 1}^\infty (1 - k^n)^{2 - d} \nonumber \\ 
& \times & \left\{ 
\exp \left[\frac{1}{2} \sum_{i} \sqrt{2 \a'}\, p_i \cdot \e_i 
\left(\frac{V''_i(0)}{V'_i(0)} \right) \right] 
\prod_{i < j} \left[ \frac{\exp \left({\cal G}(V_i(z),V_j(y)) 
\right)}{\sqrt{V'_i(z)\,V'_j(y)}}
\right]^{2\a' p_i\cdot p_j}_{z=y=0} 
\right. \nl & \times & 
\exp \left[ \sum_{i \not= j}
\left(\sqrt{2 \a'} \, p_j \cdot \e_i 
\, \left[ \partial_z  {\cal G}(V_i(z),V_j(y))
\right] _{z = y = 0} \right. \right. \nl 
& & \left. \left. \left. + \, {1 \over 2} 
\e_i \cdot \e_j \, \left[ 
\partial_z \partial_y
{\cal G}(V_i(z), V_j(y)) \right] _{z=y=0} \right) 
\right] \right\}_{\rm m.l.}~,
\label{hmastac} 
\eeqa
where the subscript ``{\rm m.l.}'' means that only terms multilinear in
all polarization vectors must be considered.

\eq{hmastac} gives the one--loop contribution to the color--ordered
scattering amplitude of $M$ gluons characterized by momenta $p_i$,
polarization vectors $\epsilon_i$ and color indices $a_i$, in the
general case in which the mass--shell ($p_i^2 = 0$) and transversality
($\epsilon_i \cdot p_i = 0$) conditions have not been
imposed\footnote{Notice that we use the metric $(- + + +)$ in
string--derived expressions.}.  Since this amplitude is color--ordered
(one needs to sum over all inequivalent orderings, to obtain the full
amplitude), the color structure is factorized and appears in the form
of a Chan--Paton factor. At one loop this factor contains at most two
traces of color matrices, corresponding to the insertion of gluons on
both boundaries of the annulus; in the planar case one has $m = 0$,
and only one trace survives, multiplied times a factor of $N = {\rm
Tr}{\bf 1}$. The string coupling $g_S$ can be related to the gauge
coupling in $d = 4 - 2 \epsilon$ dimensions by matching the values of
the tree--level three--point amplitude derived from string theory with
the corresponding Yang--Mills Feynman rule. With our choice of
normalization for the the $SU(N)$ generators in the fundamental
representation, $\lambda^a$,
\beq
{\rm Tr}(\lambda^a \, \lambda^b) = \frac{1}{2} \, \delta^{a b}~,
\label{gennorm}
\eeq
the matching condition reads \cite{big}
\beq
g_S = \frac{g_d}{2} (2 \a')^{1 - d/4}~,
\label{gsg}
\eeq
where $g_d = g_{YM} \mu^\epsilon$, and $\a'$, the Regge slope, is
inversely proportional to the string tension $T$.  In the field theory
limit one considers amplitudes where the typical energy is much
smaller than $\a'^{-1/2}$, that is $\alpha' \, p_i \cdot p_j \ll
1$. This limit can be computed by formally considering $\alpha' \to
0$.

The Koba--Nielsen variables $z_i$ in \eq{hmastac} represent the
locations of gluon insertions on the boundary of the one--loop
world--sheet.  The symbol $d V_{abc}$ in the integration measure
reminds us that projective invariance of the string amplitude can be
used to fix the location of three integration variables, chosen among
the punctures $z_i$ or the fixed points. Multipliers, on the other
hand, cannot be fixed; as we will see, they are related to Schwinger
parameters measuring the total length of the loops. In the present
case, it is convenient to fix $\xi = \infty$, $\eta = 0$; next, we fix
the position of one of the punctures, setting $z_1 = 1$, as indicated
in \fig{scho}.

The dependence of \eq{hmastac} on the projective transformations
$V_i(z)$ deserves further comment. Local coordinates around the
punctures must be introduced in the operator formalism in order to
perform the sewing procedure \cite{copgroup}, which leads to the
construction of loop amplitudes from a tree--level multi--leg vertex
operator (the `$N$--Reggeon vertex')\footnote{Roughly speaking, the
sewing procedure generates a $g$--loop, $m$--point amplitude by
inserting a propagator between two external states of a $(g -
1)$--loop, $(m + 2)$--point amplitude, and summing over the whole
spectrum of string states propagating in that channel.}.  If one
imposes the mass--shell and transversality conditions, it can be
easily verified that the $V_i$ dependence in \eq{hmastac} cancels, as
a consequence of momentum conservation. Furthermore, as we will see,
even off--shell it is possible to reabsorb all the $V_i$ dependence in
a simple redefinition of the Green function. This can be achieved by
defining
\beq
G(V_i(z), V_j(y)) = {\cal G}(V_i(z), V_j(y)) - \frac{1}{2} \log |V'_i(z)|
- \frac{1}{2} \log |V'_j(y)|~.
\label{pregcalg}
\eeq
Substituting Eqs.~(\ref{gsg}) and (\ref{pregcalg}) into \eq{hmastac},
and making use of \eq{local}, we recover a simpler expression of the
master formula, which was used in previous work \cite{big}, and which
has no explicit dependence on the projective transformations
$V_i(z)$. What is most significant is the fact that the term
proportional to $p_i \cdot \epsilon_i$ is canceled {\it without}
imposing the transversality condition to external gluons. The formula
we find is then applicable to unphysical as well as to physical
polarizations. Focusing on the planar case, the `practical version' of
the master formula can be written as
\beqa
& & \hspace{-0.5cm} A^{(1)}_P (1, \ldots, M) ~=~ N \, {\rm Tr}(\lambda^{a_M}
\cdots \lambda^{a_1}) \, \frac{g_d^M}{(4\pi)^{d/2}} \,
(2 \a')^{(M - d)/2} \int_{0}^1 \frac{dk}{k^2}~ \prod_{n = 1}^\infty
(1 - k^n)^{2 - d} \nl 
& \times &  \int_{k}^{1} \!\!\! dz_M \, \int_{z_{M}}^{1} \!\!\! dz_{M-1}
\, \ldots\! \int_{z_3}^{1} \!\!\! dz_2 \,  
\left( - \frac{\log k}{2} \right)^{- d/2} \prod_{i<j} \Big(
\exp \left[ G(z_i,z_j) \right] \Big)^{2 \a' p_i \cdot p_j} 
\label{onemaster} \\
& \times &  \left\{ \exp \left[ \sum_{i \not=j}
\left(\sqrt{2 \a'} \, p_j \cdot \e_i 
\, \partial_i  G(z_i, z_j)+ {1 \over 2} 
\, \e_i \cdot \e_j \, \partial_i \partial_j
G(z_i, z_j) \right) \right] \right\}_{\rm m.l.}~.
\nonumber
\eeqa
To extract results from \eq{onemaster} in the case of off--shell
gluons, one must of course make a choice for the $V_i (z)$. This
choice can be guided by two--dimensional geometry. It is known in fact
that, for any Riemann surface, the Green function contains the
logarithm of the prime form, which is essentially a well defined
analytic generalization of the monomial $z_i - z_j$ to higher--genus
surfaces. The prime form is not a function on the surface, but rather
a differential form of weight $\{-1/2, -1/2\}$. One sees from
\eq{pregcalg} that one can construct a well behaved Green function on
the surface by choosing the $V_i(z)$'s so that their derivatives
$V'_i(0)$ behave as differential forms of weight $-1$. The only
globally defined one--forms on a genus $g$ Riemann surface are the $g$
abelian differentials, so, at one loop, this criterion leads to an
essentially unique choice.  One must choose
\beq
V_i' (0) = \left( \omega (z_i) \right)^{-1}~,
\label{vpri}
\eeq
where $\omega(z)$ is related to the unique abelian differential on the torus,
which for our choice of fixed points ($\xi=\infty,\,\eta=0$) is simply
$d z/z$. We are lead uniquely to the choice $V_i'(0) =
z_i$, already adopted in Refs.~\cite{big,scal}. With this choice, the
modified open string Green function $G(z_i, z_j)$ defined in
\eq{pregcalg} is given by
\beqa
G (z_i, z_j) & = & \log \left( \left| \sqrt{\frac{z_i}{z_j}} - 
\sqrt{\frac{z_j}{z_i}} \right| \right) 
+ \frac{1}{2 \log k} \left( \log \frac{z_i}{z_j} \right)^2 \nl
& + &
\log \left[ \prod_{n = 1}^\infty \frac{\left(1 - k^n \frac{z_j}{z_i} \right)
\left(1 - k^n \frac{z_i}{z_j} \right)}{\left(1 - k^n \right)^2}
\right]~;
\label{onegreen}
\eeqa
notice that $G(z_i, z_j)$ is actually a function only of the ratio
$\rho_{i j} = z_i/z_j$, so that it is translationally invariant on the
torus parametrized by coordinates $\nu_i \sim \log(z_i)$.

We conclude this section with a comment on the integration region over
the moduli given in \eq{onemaster}.  Notice that the $z_i$'s are
ordered between $k$ and $z_1 = 1$. In fact, in the Schottky
parametrization, for a planar configuration, all punctures are
constrained to lie on the same boundary of the string world--sheet,
thus, having fixed $z_1 = 1$, all other $z_i$ should be integrated
over the interval $ B = \sqrt{k} < z_i < B' = 1/\sqrt{k}$, with the
restriction on the ordering implied by the color trace.  This would
complicate the calculation of the field theory limit, since there
would be contributions both from $z_i \to \sqrt{k} \to 0$ and from
$z_i \to 1/\sqrt{k} \to \infty$.  It is possible to bypass this
practical difficulty by making use of the fact that the string
integrand is modular invariant, which in particular implies that the
interval $[1,1/\sqrt{k}]$ can be mapped onto the interval
$[k,\sqrt{k}]$. One can easily show that the effective one--loop Green
function in \eq{pregcalg} satisfies
\beqa
G \left(\rho_{j i}; k\right) & = & G \left(\rho_{i j}; k\right) \nl 
G \left( k \rho_{j i }; k\right) & = & G \left(\rho_{i j}; k\right)~~,
\label{modg1}
\eeqa
where $\rho_{ij} = z_i/z_j$ and with a slight abuse of notation we write
$G \left(\rho_{ij}; k \right) \equiv G \left(z_j, z_i \right)$. 
Using these properties, one can map all configurations with a subset of 
punctures in the interval $[1,1/\sqrt{k}]$ to configurations in which those
punctures have been moved to the interval $[k,\sqrt{k}]$, preserving the
ordering on the circle. This procedure yields the integration region in
\eq{onemaster}.

\sect{General structure of the field theory limit}
\label{GeneralFT}

In the formal limit $\a' \to 0$, \eq{onemaster} gives a vanishing or a
divergent result, depending on the number of external gluons, $M$. It
is clear that a finite result can be obtained only by considering
singular regions of integration, and parametrizing the singularities
in order to generate powers of $\a'$, to compensate the explicit
factor of $\a'^{(M-d)/2}$.  In order to do this, it will be necessary
to change integration variables from the dimensionless quantities
$\{k, z_i\}$ to dimensionful parameters characterizing the size and
shape of the string world--sheet in units of $\a'$.

The appropriate change of variables is once again suggested by the
geometric structure emerging from the operator formalism. The sewing
procedure, in fact, indicates that when the integrand of
\eq{onemaster} is written out as a Laurent series expansion in powers
of the multiplier $k$, the coefficient of $k^n$ represents the
contribution to the amplitude due to the circulation in the loop of
states belonging to mass level $n+1$.  Since the integrand has no
power--law divergence as $k \to 0$, except the explicit factor of
$k^{-2}$, one can conclude that the terms proportional to $k^{-2}$
correspond to tachyon propagation (mass level $n = -1$), while the
contribution of gluons (mass level $n = 0$) will be given by terms
proportional to $k^{-1}$. This interpretation of the role played by
$k$ also rules out the possibility of expressing $k$ directly as the
ratio of a dimensionful parameter and $\a'$; furthermore, after
isolating the correct power of $k$, the only singularities left in the
integrand are logarithmic, and the measure of integration for the
multiplier is just $d \log k$. This suggests \cite{kaj} that one
should identify the logarithm of the multiplier with the length of the
string loop, measured in units of $\a'$, in agreement with the fact
that the limit $k \to 0$ corresponds to a very long and narrow
annulus, as seen explicitly in Fig.~\ref{scho}.  It is then natural to
think of the puncture coordinates as associated to fractional lengths
in a similar way. This leads to the choice
\beq
\log k = - \frac{t}{\a'}~, \qquad \qquad \log z_i = - \frac{t_i}{\a'}~,
\label{proptimes}
\eeq
which has been employed in the past for both gluon \cite{beko,big}
and scalar \cite{scal} amplitudes. It turns out that the variables
$t_i$ and $t$ play the role of Schwinger parameters associated to
combinations of field theory propagators, as we will see explicitly 
below.

Once the change of variables in \eq{proptimes} has been effected, the
field theory limit can be reached by taking $\a' \to 0$ with $\{t,
t_i\}$ fixed. Thus one expands the integrand of \eq{onemaster} in
powers of $k$ and $z_i$, retains the coefficients of $k^{-1}$ and
$z_i^{-1}$, and keeps all terms that have a finite limit as $\a' \to
0$. As we will see, one must be careful to include all sources of
$\a'$ dependence, and all singular regions of integration.

This is a representative case of the general strategy that must be
employed also for amplitudes with more loops: the zero--slope limit
has to be taken after introducing the dimensionful field theory
variables that have to be kept fixed. The exact form of the change of
integration variables identifies the particular corner of moduli space
being considered.  To get the complete result, it is necessary to make
sure that all singular corners of integration have been taken into
account. In particular, in the case of one--loop gluon amplitudes, it
is immediately clear that the region identified by \eq{proptimes}
cannot give the complete answer.  Consider in fact the counting of
powers of $\a'$ in \eq{onemaster} after implementing \eq{proptimes}:
one finds an overall factor of $(\a'/2)^{- M/2}$ from the
normalization and the change of variables, and a further factor of
$\a'^m$, with $0 \leq m \leq M/2$, from the terms multilinear in the
polarization vectors arising from the expansion of the second
exponential.  Specifically, only the terms containing no double
derivatives of the Green function display a complete cancellation of
the overall power of $\a'$, giving immediately a contribution to the
field theory limit. All terms containing double derivatives of $G$
still have an overall {\it negative} power of $\a'$, and one must find
further sources of positive powers of $\a'$ in order to generate
finite contributions from these terms.  There are two such sources,
which generate classes of contributions which are easily identified in
the field theory limit.

\begin{itemize}

\item Positive powers of $\a'$ will arise from regions of integration
in which certain Schwinger parameters are small, in fact are
themselves ${\cal O}(\a')$.  Notice that so long as, say, $t_i < t_j$,
in the $\a' \to 0$ limit the punctures are strongly ordered on the
real axis, {\it i.e.} $z_j \ll z_i$.  Taking $t_i - t_j = {\cal
O}(\a')$, on the other hand, means that the punctures remain close to
each other as the field theory limit is taken. These integration
regions must then, and do, correspond to four--point vertices in field
theory. From the point of view of string theory such contributions
only arise naturally in gluon amplitudes: if one takes a field theory
limit corresponding to scalars with a cubic coupling, these regions
are always suppressed by powers of $\a'$. Notice also that the
counting of powers of $\a'$ described above correctly reproduces the
counting of four--point vertices in gluon diagrams: in an $M$--point
one--loop amplitude there can be at most $M/2$ four--point vertices
(or $(M-1)/2$, for odd $M$), which is also the maximum number of
powers of $\a'$ that can be extracted with this method in
\eq{onemaster} without getting a vanishing result. For obvious
dimensional reasons, diagrams with only four--point vertices have no
powers of momenta in the numerator, and thus give results in which all
polarization vectors are dotted into each other; this is also directly
evident in \eq{onemaster}.

\item A second source of positive powers of $\a'$ is the first
exponential factor of \eq{onemaster}. In fact, upon substituting
\eq{proptimes}, one easily sees that this factor takes the form
\beq
\prod_{i<j} \left[ 
\exp \left(G(z_i,z_j) \right)
\right]^{2 \a' p_i \cdot p_j} = \exp \left(c_0(t_i) + \a' c_1(t_i) 
\right)~.
\label{expalp}
\eeq
Clearly, if the overall power of $\a'$ in \eq{onemaster} is negative,
\eq{expalp} must be expanded as
\beq
\prod_{i<j} \left[ 
\exp \left(G(z_i,z_j) \right)
\right]^{2 \a' p_i \cdot p_j} =
\exp \left(c_0(t_i) \right) \left(1 + \a' c_1(t_i) + \frac{1}{2} \a'^2
c_1(t_i)^2 + \ldots \right)~,
\label{expexp}
\eeq
and thus participates in the counting of $\a'$. All terms arising from
diagrams containing only cubic vertices but containing factors of the
form $p_i \cdot p_j \ldots \epsilon_k \cdot \epsilon_l$ are generated
by string theory through this expansion.

\end{itemize}

The discussion developed so far is based upon the idea of expanding
the integrand in powers of $k$ and $z_i$, assuming that the punctures
are not too close to each other, so that the Green function is not
dominated by its short distance singularity as $z_i \to z_j$. As a
result, one gets all 1PI contributions to the amplitude. As pointed
out long ago \cite{beko}, one can also get one--particle--reducible
contribution, by considering the opposite limit, taking $z_i \to z_j$
{\it before} the field theory limit, driven by $k \to 0$. This
procedure is also a source of powers of $\a'$, since it generates
dimensionless gluon propagators in the chosen channel, in the form
$1/(\a' p_i \cdot p_j)$. This limit will be discussed in more detail
in \secn{Reg3}.

Summarizing, we have tentatively identified all the regions of
integration in moduli space contributing to the field theory limit. We
see that in this limit the string amplitude decomposes into a sum of
contributions associated to graphs that, as we shall see, are in
straightforward correspondence with the Feynman graphs of the limiting
field theory. Specifically, the graphs for a given amplitude can be
enumerated by counting the Feynman graphs with the same number of
external legs in a scalar theory with both cubic and quartic
interactions. Each string--derived graph corresponds to a sum of
Feynman graphs in Yang--Mills theory, involving the propagation of
both ghosts and gluons. Furthermore, the field theory limit yields
directly the Schwinger parameter integral, having bypassed all Lorentz
and color algebra, loop momentum integration, and the decomposition of
tensor integrals into scalar building blocks \cite{pv,vanver}.  We
will refer to the string--derived graphs as `scalar graphs'. Scalar
graphs are directly expressed in terms of scalar integrals, albeit
with a numerator structure reminiscent of the presence of momenta in
Yang--Mills vertices. At least at one loop, string theory effectively
succeeds in reducing the complexity of the calculation of a
Yang--Mills amplitude to a level close to a scalar amplitude, even in
the general case of off--shell gluons with arbitrary polarizations.

In the next sections, we will describe in more detail how different
scalar graphs must be computed starting from \eq{onemaster}. We will
consider in turn irreducible graphs with only cubic vertices, arising
from the region in which the punctures are strongly ordered
(\secn{Reg1}), then irreducible graphs with quartic vertices, arising
when the strong ordering condition is relaxed (\secn{Reg2}), and
finally reducible graphs (\secn{Reg3}). In each case we will give
concrete examples.  Because the graphs are computed with off--shell
gluons, their values are gauge--dependent in field theory.  One is
then able to make a precise identification of the gauge choice made by
string theory, and in fact our calculations confirm the results of
Refs.~\cite{bedu} and \cite{big}: the string computes the amplitudes
in a combination of the background Feynman gauge (for irreducible
diagrams) with the Gervais--Neveu gauge (for trees attached to the
loop). It is tempting to identify the dependence of the off--shell
string amplitude on the choice of projective coordinates $V_i'(0)$
with a gauge dependence in the field theory limit. However we have so
far been unable to make this identification precise.

We conclude this section by listing the explicit expressions that we
will need for the Green function and its derivatives in the various
relevant limits.  In the limit $k \to 0$, \eq{onegreen} and its
derivatives have the expansions
\beqa
G \left(\rho_{ij}; k \right) & = & \log (1 - \rho_{ij}) - \frac{1}{2}
\log \rho_{ij} + \frac{\log^2 \rho_{ij}}{2 \log k} - k \left(\rho_{ij}
+ \frac{1}{\rho_{ij}} - 2 \right) + {\cal O} (k^2)~, \nl
\partial_i G \left(\rho_{ij}, k \right) & = & \frac{1}{z_j} 
\left[- \frac{1}{1 - \rho_{ij} } - \frac{1}{2 \rho_{ij} } + \frac{\log
\rho_{ij}}{\rho_{ij} \log k} - k \left(- \frac{1}{\rho_{ij} ^2} + 1
\right) \right] + {\cal O}(k^2)~, \nl
%new derivative
\partial_j G \left(\rho_{ij}; k \right) & = & -\frac{z_i}{z_j^2} 
\left[- \frac{1}{1 - \rho_{ij} } - \frac{1}{2 \rho_{ij} } + \frac{\log
\rho_{ij}}{\rho_{ij} \log k} - k \left(- \frac{1}{\rho_{ij} ^2} + 1
\right) \right] + {\cal O}(k^2)~, \nl
\partial_i \partial_j G \left(\rho_{ij}; k \right) & = & \frac{1}{z_i
z_j} \left[ - \frac{1}{\log k} + \frac{\rho_{ij}}{(1 - \rho_{ij} )^2}
+ k \left(\frac{1}{\rho_{ij}} + \rho_{ij} \right) \right] + {\cal
O}(k^2)~, 
\label{expgr}
\eeqa
where $\rho_{ij} = z_i/z_j < 1$, since the punctures are radially
ordered on the boundary.

As we will see in \secn{Reg1}, if the two punctures are on different
boundaries, so that their coordinates have opposite sign in Schottky
parametrization, the Green function is slightly modified and becomes
\beq
\widehat{G} \left(\rho_{ij}, k \right) = \log (1 + |\rho_{ij}|) -
\frac{1}{2} \log |\rho_{ij}| + \frac{\log^2 |\rho_{ij}|}{2 \log k} +
k(|\rho_{ij}| + \frac{1}{|\rho_{ij}|} + 2) + {\cal O} (k^2)~.
\label{gnp}
\eeq
When considering reducible diagrams, we will also need the limit of
the Green function as $\rho \to 1$. In this limit, the logarithmic
singularity dominates, and one can use simply
\beq
G \left(\rho; k \right) = \log (1 - \rho) + {\cal O}(1 - \rho)~.
\label{pinchg}
\eeq
When the two punctures are on different boundaries, there is no
logarithmic singularity in the integration domain, at least when
considering three--gluon vertices, so the corresponding limit for the
non--planar Green function is not needed. We are now ready to turn to 
a more detailed analysis of the field theory limit for different 
scalar graph topologies.

\sect{Irreducible diagrams with cubic vertices}
\label{Reg1}

Irreducible scalar diagrams with only cubic vertices are the most
straightforward to compute, since they arise naturally from the region
of moduli space parametrized by \eq{proptimes} in the limit $\a' \to
0$.  As pointed out above, in this region each gluon insertion on the
annulus is widely separated from the contiguous ones: thus, on both
sides of each external state there are loop propagators.  Consider,
for example, two contiguous punctures $z_i$ and $z_{i + 1}$.
According to \eq{proptimes}, in the field theory limit we set
\beq
z_i = e^{- t_i/\a'} \qquad ; \qquad z_{i + 1} = e^{- t_{i + 1}/\a'}~.
\label{ex1}
\eeq
The existence of a loop propagator between the two insertions,
requires that its proper time, {\it i.e.} $t_{i+1} - t_i$, be kept finite 
as $\a' \to 0$. Then 
\beq
\frac{t_{i + 1} - t_i}{\a'} \to \infty \quad \Rightarrow \quad
\frac{z_{i + 1}}{z_i} \to 0 \quad \Rightarrow \quad z_{i + 1} \ll z_i~,
\eeq
thus we conclude that the gluon labeled by $i$ connects to the loop
through a cubic vertex if the Koba-Nielsen variables adjacent to it
are strongly ordered, $z_{i + 1} \ll z_i \ll z_{i - 1}$\footnote{We
work in the configuration $0 \leq k \leq z_M \leq \ldots \leq z_1 =
1$. To state that $z_M$ is widely separated from $z_1 = 1$, we may
simply require it to be widely separated from $k$, since points $k$
and 1 are identified on the annulus.}.

%%%%%%%%%%%%%%%%%%%%%%%%%%%%%%%%%%%%%%%%%%%%%%%%%%%%%%%%%%%%%%%%%%%%%%%%%%
\begin{figure}
\begin{center}
\begin{picture}(370,150)(40,-15)
\GlueArc(100,60)(30,-90,45){3}{5}
\GlueArc(100,60)(30,45,135){3}{3}
\LongArrowArcn(100,60)(20,135,45)
\Text(100,70)[]{$k$} 
\GlueArc(100,60)(30,135,270){3}{5}
\Gluon(58,103)(78,81){3}{2}
\Gluon(122,81)(142,102){3}{2}
\Gluon(100,30)(100,2){3}{2}
\Text(100,100)[]{$x_1$}
\Text(58,60)[]{$x_3 $}
\Text(142,60)[]{$x_2 $}
\Text(60,82)[b]{$p_1$}
\Text(142,82)[b]{$p_2 $}
\Text(90,10)[]{$p_3 $}
\Text(53,108)[]{A}
\Oval(53,108)(7,7)(0)
\Text(147,107)[]{A}
\Oval(147,107)(7,7)(0)
\Text(100,-5)[]{A}
\Oval(100,-5)(7,7)(0)
\DashArrowArc(230,60)(30,-90,45){3}
\DashArrowArc(230,60)(30,45,135){3}
\DashArrowArc(230,60)(30,135,270){3}
\Gluon(188,103)(208,81){3}{2}
\Gluon(252,81)(272,102){3}{2}
\Gluon(230,30)(230,2){3}{2}
\Text(183,108)[]{A}
\Oval(183,108)(7,7)(0)
\Text(277,107)[]{A}
\Oval(277,107)(7,7)(0)
\Text(230,-5)[]{A}
\Oval(230,-5)(7,7)(0)
\DashArrowArcn(360,60)(30,45,-90){3}
\DashArrowArcn(360,60)(30,135,45){3}
\DashArrowArcn(360,60)(30,270,135){3}
\Gluon(318,103)(338,81){3}{2}
\Gluon(382,81)(402,102){3}{2}
\Gluon(360,30)(360,2){3}{2}
\Text(313,108)[]{A}
\Oval(313,108)(7,7)(0)
\Text(407,107)[]{A}
\Oval(407,107)(7,7)(0)
\Text(360,-5)[]{A}
\Oval(360,-5)(7,7)(0)
\end{picture}
\end{center}
\caption{Three--point diagrams: external gluons are background fields,
 marked by an A. Momenta $p_i$ are incoming.}\label{threep}
\end{figure}
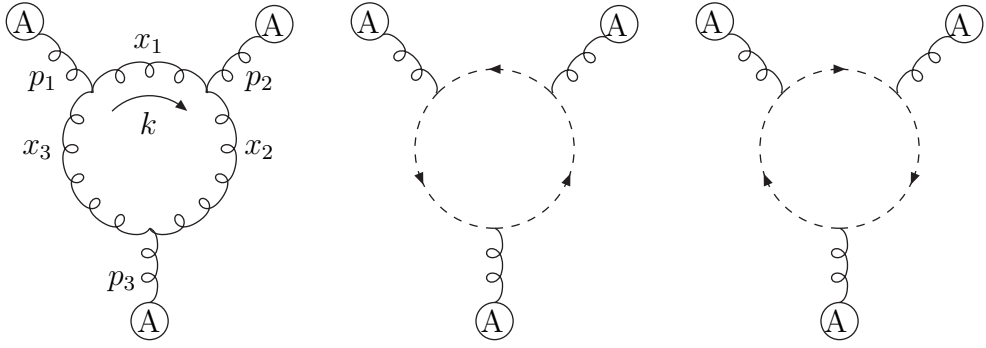
%%%%%%%%%%%%%%%%%%%%%%%%%%%%%%%%%%%%%%%%%%%%%%%%%%%%%%%%%%%%%%%%%%%%%

To illustrate the technique, we will begin by computing the sum of the
three diagrams given in Fig.~\ref{threep}, using \eq{onemaster} for $M
= 3$, and changing variables according to
\eq{proptimes}. Concentrating on the color ordering (3-2-1), and
retaining only the leading terms in the expansion in powers of the
multiplier, \eq{onemaster} for $M = 3$ reads
\beqa
& & A^{(1)} (1, 2, 3) = N ~\Tr (\l^3 \l^2 \l^1) 
\frac{g_d^3}{(4 \pi )^{d/2}} (2\a')^{2 - d/2} \int_0^1 \frac{dk}{k^2} 
\int_k^1 d z_3 \int_{z_3}^1 d z_2  \label{3master} \\
& \times  & \left( -\frac{\log k}{2} \right)^{- d/2} 
\left(1 + k ~(d - 2) \right) ~\exp 
\Big[ 2 \a' \left( p_1 \cdot p_2 G_{1 2} + p_1 \cdot p_3 G_{1 3} +
p_2 \cdot p_3 G_{2 3}\right) \Big] \nl
& \times & \Bigg\{ 2 \a' \left(\e_1 \cdot p_2 \partial_1 G_{2 1} + 
\e_1 \cdot p_3 \partial_1 G_{3 1} \right) 
\left( \e_2 \cdot p_3 \partial_2 G_{3 2} + \e_2 \cdot p_1 \partial_2 G_{1 2} 
\right) \nl & & \qquad \times ~\left( \e_3 \cdot p_1 \partial_3 
G_{1 3} + \e_3 \cdot p_2 \partial_3 G_{2 3} \right) \nl
& & + ~\Big[ \e_1 \cdot \e_2 \partial_1 \partial_2 G_{1 2} 
\left( \e_3 \cdot p_1 \partial_3 G_{1 3} + \e_3 \cdot p_2 \partial_3 G_{2 3} 
\right) + \left( {\it cycl.} \right) \Big] \Bigg\}~. \nonumber
\eeqa
Here $G_{i j}$ denotes the expansion of $G(\rho_{i j}, k)$ in powers
of $k$, \eq{expgr}.

To proceed, we note that changing variables to Schwinger parameters
the overall power of $\a'$ becomes $\a'^{-1}$. Thus for the terms in
\eq{3master} that do not contain double derivatives of the Green
function all polynomial dependence on $\a'$ cancels and the field
theory limit can be taken directly.  The terms proportional to double
derivatives of $G$, however, need an extra positive power of
$\a'$. This can be supplied by the term proportional to $1/\log k$ in
\eq{expgr}, but to get the complete result for these terms it is
necessary to expand the exponential as well, according to
\eq{expexp}. Once this is done, the field theory limit can be taken by
picking the coefficient of $\a'^0 k^0 z_i^0$ in the resulting
polynomial, while the exponential factor becomes simply
\beqa
& & \exp \Big[ 2 \a' \left( p_1 \cdot p_2 G_{1 2} + p_1 \cdot p_3 G_{1 3} +
p_2 \cdot p_3 G_{2 3}\right) \Big] \to \label{3gr} \\
& & \hspace{-4pt}
\exp \left[  p_1 \cdot p_2 t_2 \left(1 - \frac{t_2}{t} \right)
+ p_1 \cdot p_3 t_3 \left(1 - \frac{t_3}{t} \right) 
+ p_2 \cdot p_3 \left(t_3 - t_2\right) \left(1 - \frac{t_3 - t_2}{t} \right)
\right]~. \nonumber
\eeqa

To write our final answer in the most symmetric manner, it is
convenient to introduce the dimensionless Feynman parameters $x_1 =
t_2/t$, $x_2 = (t_3 - t_2)/t$, $x_3 = 1 - x_1 - x_2$, and to define
$x_{ij}$ as the sum of all Feynman parameters found between gluons $i$
and $j$, when moving clockwise around the loop (see \fig{threep}), so
that, for example, $x_{13} = x_1 + x_2$. The complete expression for
\eq{3master} in the field theory limit becomes then
\beqa
A^{(1)} (p_1, p_2, p_3) & = &  N ~\Tr (\l^3 \l^2 \l^1) ~\frac{g_d^3}{(4 
\pi )^{d/2}} \int_0^\infty d t ~t^{2 - d/2} \nl
& \times & \int_0^1 d x_1~d x_2~d x_3 ~\delta (1-\sum_{j=1}^3 x_j) \nl 
& \times & \exp \left[ - t \left(x_1 x_2 p_2^2 + x_2 x_3 p_3^2 + 
x_3 x_1 p_1^2 \right) \right] \nl
& \times & \Bigg\{ - \sum_{i=2,3;j=1,3;k=1,2} \e_1 \cdot p_i \e_2 
\cdot p_j \e_3 \cdot p_k \nl
& & ~~\Big[(d - 2) (x_{1i} - x_{i1})(x_{2j} - x_{j2})(x_{3k} - x_{k3}) \nl
& & ~~+ ~8 (x_{1i} - x_{i1} + x_{2j} - x_{j2} + x_{3k} - x_{k3})
\Big] \nl
& & ~~+ \sum_{(i,j,k)=(1,2,3),(2,3,1),(3,1,2)}
\e_i \cdot \e_j \e_k \cdot p_i \label{full3} \\
& & \left[ \frac{2 ( d - 2)}{t}(x_{ik} - x_{ki})+ 16 x_{ki} p_i \cdot p_j
+ 8 p_j^2 \right]  \nl
& & ~~+ \sum_{(i,j,k)=(1,2,3),(2,3,1),(3,1,2)} \e_i \cdot \e_j \e_k 
\cdot p_j \nl 
& & \left[ \frac{2 (d - 2)}{t}(x_{jk} - x_{kj}) + 16 x_{kj} 
p_i \cdot p_j + 8 p_j^2 - 8 p_k^2 \right] \Bigg\}~. \nonumber   
\eeqa
In writing \eq{full3}, we kept all terms proportional to
longitudinal gluon polarizations, but to get a symmetric expression we
replaced, say, $\epsilon_1 \cdot p_1$ with $ - \epsilon_1 \cdot
p_2 - \epsilon_1 \cdot p_3$. The result exactly matches the
contribution of the corresponding color ordering to the three diagrams
in \fig{threep}, computed using the background field method, with the
Feynman rules given for example in \cite{abbott}\footnote{The string
result must be multiplied by a factor $i$, since the string
rules give directly the $T$ matrix.}. We note in passing that the
function appearing in the exponent of \eq{full3} depends only on the 
topology of the diagram, and thus is the same that would appear in a 
scalar theory.

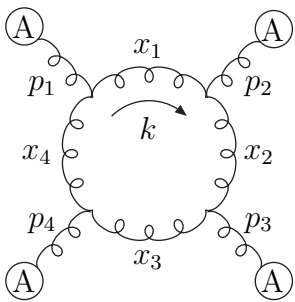
\begin{figure}
\begin{center}
\begin{picture}(200,150)(0,-15)
\GlueArc(100,60)(30,-45,45){3}{3}
\GlueArc(100,60)(30,45,135){3}{3}
\GlueArc(100,60)(30,135,225){3}{3}
\GlueArc(100,60)(30,225,315){3}{3}
\LongArrowArcn(100,60)(20,135,45)
\Text(100,70)[]{$k$} 
\Gluon(58,103)(78,81){3}{2} \Gluon(58,17)(78,38){3}{2}
\Gluon(122,81)(142,102){3}{2} \Gluon(122,38)(142,17){3}{2}
\Text(100,100)[]{$x_1$}
\Text(58,60)[]{$x_4 $}
\Text(142,60)[]{$x_2 $}
\Text(100,20)[]{$x_3 $}
\Text(60,82)[b]{$p_1$}
\Text(142,82)[b]{$p_2 $}
\Text(142,34)[]{$p_3 $}
\Text(60,34)[]{$p_4 $}
\Text(53,108)[]{A}
\Oval(53,108)(7,7)(0)
\Text(53,12)[]{A}
\Oval(53,12)(7,7)(0)
\Text(147,107)[]{A}
\Oval(147,107)(7,7)(0)
\Text(147,12)[]{A}
\Oval(147,12)(7,7)(0)
\end{picture}
\end{center}
\caption{Four--point diagram: external gluons are background fields,
 marked by an A. Momenta $p_i$ are incoming.}\label{fourp}
\end{figure}

Using symbolic algebraic manipulation programs, one can check at the
level of integrands that our prescription gives the correct results
for scalar graphs also for amplitudes with more gluons. We have tested
the irreducible graph with only cubic vertices in the case of the four
gluon amplitude, but the complete result is too lengthy to report
here. It is more interesting to use the four--point amplitude as a
testing ground for the computation of non--planar contributions, since
the first non--trivial example arises in this case. Consider then the
scalar graph containing the four--gluon diagram in \fig{fourp}, taking
for simplicity on--shell gluons.
%%%%%%%%%%%%%%%%%%%%%%%%%%%%%%%%%%%%%%%%%%%%%%%%%%%%%%%%%%%%%%%%%%%%%
%%%%%%%%%%%%%%%%%%%%%%%%%%%%%%%%%%%%%%%%%%%%%%%%%%%%%%%%%%%%%%%%%%%%%

With a growing number of external particles, it is convenient to fix
helicities, which helps to reduce the number of non--vanishing Lorentz
invariant combinations of external momenta and polarizations.  Using
standard helicity techniques, we may choose for example the
configuration $\e^+(p_1,p_4)$, $\e^+(p_2,p_4)$, $\e^-(p_3,p_1)$,
$\e^-(p_4,p_1)$, where the first argument is the external gluon
momentum, while the second is the reference momentum, as explained in
\cite{manpa}.  In this case the only five non--vanishing scalar
products are $\e_2 \cdot \e_3$, $\e_1 \cdot p_3 = -\e_1 \cdot p_2$,
$\e_2 \cdot p_1 = -\e_2 \cdot p_3$, $\e_3 \cdot p_4 = -\e_3 \cdot
p_2$, and $\e_4 \cdot p_2=-\e_4 \cdot p_3$.  Summing gluon and ghost
loops, we find that the scalar graph, in the background field method,
has the expression
\beqa
A^{(1)} (p_1, p_2, p_3, p_4) & = & 8 ~i \, {\cal C} \frac{g_d^4}{(4 
\pi )^{d/2}} \int_0^{\infty } \! d t~t^{3 - d/2} \int_0^1 d x_1 d x_2 d x_3 
d x_4 ~\delta (1-\sum_{j=1}^4 x_j)\nl
& \times & \exp \left[t \left(s_{12} x_2 x_4 + s_{23} x_1 x_3 \right) \right]~
(\e_1 \cdot p_2)(\e_4 \cdot p_3) \\
& \times & \Bigg\{ 2  (\e_2 \cdot p_3) (\e_3 \cdot p_2) \Big[ 
- 2 ~ x_2 x_4 + (d - 2)  x_2^2 (x_3 + x_4)(x_1 + x_4) \Big] \nl
& & ~~+ (\e_2 \cdot \e_3) \Big[  s_{23} \left( (x_1 + x_3 + x_4)^2 + x_2^2
\right) + (d - 2)\frac{1}{t}  x_2^2 \Big] \Bigg\}~, \nonumber
\label{sun4}
\eeqa
where the color factor is 
\beqa
{\cal C} & = & N \left[ \Tr(\l^1 \l^2 \l^3 \l^4) + \Tr(\l^4 \l^3 \l^2 \l^1)
\right] \nl
& + & 2 ~\left[\Tr(\l^1 \l^2) \Tr(\l^3 \l^4) + \Tr(\l^1 \l^3) \Tr(\l^2 \l^4)
+ \Tr(\l^1 \l^4) \Tr(\l^2 \l^3)\right]~.
\label{c4}
\eeqa 
Leading color terms can be obtained from the planar string amplitude,
exactly as in the previous example. Here we focus on subleading terms
proportional to double traces, which are reproduced by non--planar
string amplitudes. They were absent in the three--point amplitude,
since $\Tr(\l^i) = 0$.

Consider, for example, the term proportional to $\Tr(\l^1 \l^2)
\Tr(\l^3 \l^4)$. The string amplitude with this Chan--Paton factor is
the one with $z_ 1$ and $z_2$ on one boundary (say, the positive real
axis in the Schottky representation, Section~\ref{General}, with $z_1
= 1$) while $z_3$ and $z_4$ are on the other boundary.  The
non--planar master formula is similar to the planar one, with small
but significant differences.  First of all, when a Green function
connects two punctures on different boundaries, its non--planar
version, \eq{gnp}, must be used. Also, the correct expression for the
projective transformations is $V'_i(0) = |z_i|$.  Finally, $z_2$ and
$z_3$ are not ordered, so that the resulting integration regions is
the interval $[-1, -k]$ for both punctures.  To compare non--planar
and planar contributions, we change variables according to $z_3 \to -
z_3'$ and $z_4 \to - z_4'$; all punctures are now integrated on the
positive real axis, just as in the planar case, however now it is
necessary to distinguish six different orderings of the four
punctures, namely
\begin{center}
1) $k\ll z_4'\ll z_3'\ll z_2 \ll 1$ \\
2) $k\ll z_2 \ll z_3'\ll z_4'\ll 1$ \\
3) $k\ll z_3'\ll z_2 \ll z_4'\ll 1$ \\
4) $k\ll z_4'\ll z_2 \ll z_3'\ll 1$ \\
5) $k\ll z_3'\ll z_4'\ll z_2 \ll 1$ \\
6) $k\ll z_2 \ll z_4'\ll z_3'\ll 1$
\end{center}
Regions $1$ and $2$ correspond to the cyclic orderings $(4-3-2-1)$ and
$(1-2-3-4)$, respectively. The resulting integrands, expressed in
terms of $z_2$, $z_3'$ and $z_4'$, are equal, but, at the string
level, they differ from the planar amplitude. The difference, however,
vanishes in the field theory limit, so the two regions add up to give
the relative factor of $2$ between the single trace planar
contribution proportional to $N \Tr(\l^1\l^2\l^3\l^4)$ and the double
trace contribution proportional to $\Tr(\l^1\l^2)\Tr(\l^3\l^4)$ in
\eq{c4}.

Regions $3$, $4$, $5$ and $6$ correspond to the other cyclic orderings
($(1-3-2-4)$, $(1-4-2-3)$, $(1-3-4-2)$, $(1-2-4-3)$ respectively), so
they are to be identified with the subleading terms of the other
color--ordered field theory diagrams.  This same mechanism has been
observed for scalar diagrams \cite{scal}. The fact that the
differences between planar and non--planar integrands vanish in the
field theory limit is responsible for the relations between single
trace and double trace subamplitudes, noticed already in
\cite{bekoco}.

\sect{Irreducible diagrams with quartic vertices}
\label{Reg2}

Quartic vertices arise in the field theory limit when the strong
ordering condition between two punctures is relaxed. Changes of
variables such as the one given in \eq{ex1}, in fact, generate a
certain number of explicit powers of $\a'$ in the integration
measure. It is clear however, since Schwinger parameters are
dimensionful, that there are potential ${\cal O} (\a')$ corrections
arising from regions of integration where the Schwinger parameters
themselves, or their differences, are ${\cal O} (\a')$. Such
corrections are negligible in a scalar theory, where all terms in the
integrand have a uniform power of $\a'$ when expressed in terms of the
$t_i$'s; in a gauge theory amplitude, however, these terms do
contribute, as explained in \secn{GeneralFT}.

Suppose we want two consecutive punctures, say $z_i$ and $z_{i + 1}$,
to attach to the loop through a quartic vertex: the propagator between
them would have a Schwinger parameter proportional to $t_{i + 1} -
t_i$, so we must impose that this vanish in the zero--slope limit,
setting
\beq
t_{i + 1} - t_i = {\cal O} (\a') \quad \Rightarrow \quad
\frac{z_{i + 1}}{z_i} \equiv y = {\cal O} (1)~.
\label{regII}
\eeq
The correct procedure to take into account this region of integration
is to change variables according to, say, $z_i = \exp (-t_i/\a')$, and
$z_{i + 1} = y_i z_i$, subsequently integrating over $y_i$ in a
neighborhood of $y_i = 1$. Notice that we now have a different overall
power of $\a'$, since $y_i$ is independent of $\a'$, so we will be
considering a different set of terms in the expansion of the integrand
in powers of $\a'$.

To define precisely the $y_i$ integration, consider, for example, the 
integration region for the puncture $z_2$,
\beq 
\int_{z_3}^1 \frac{dz_2}{z_2} \to \frac{1}{\a'} 
\int_0^{t_3} d t_2~.
\label{zintr}
\eeq
If we relax the condition of strong ordering of the punctures, we must
explicitly consider regions in which $t_2$ or $t_3 - t_2$ are ${\cal O}
(\a')$. In these regions the counting of powers of $\a'$ is clearly
different, as can be seen by writing
\beq
\frac{1}{\a'} \int_0^{t_3} d t_2 =
\frac{1}{\a'} \int_0^{b \a'} d t_2 +
\frac{1}{\a'} \int_{b \a'}^{t_3 - c \a'} d t_2 +
\frac{1}{\a'} \int_{t_3 - c \a'}^{t_3} d t_2~,
\label{split}
\eeq
where $b$ and $c$ are arbitrary ${\cal O} (1)$ constants.  The
contribution of diagrams with three--point vertices, discussed in the
previous section, arises from integrals such as the second term on the
RHS of \eq{split}, where we pick the terms in the integrand that
cancel the explicit factor of $1/\a'$, and thus we may neglect the
$\a'$ dependence in the limits of integration. The other two integrals
in \eq{split} do not have an overall power of $\a'$, because the size
of the integration region vanishes with $\a'$. Defining $t_2 = \a'
\tau_1$ in the first integral, and $t_3 - t_2 = \a' \tau_2$ in the
second integral (corresponding to $y_i = \exp ( - \tau_i)$) we obtain
\beq
\frac{1}{\a'} \int_0^{t_3} d t_2 =
\int_{0}^{b} d \tau_1 +
\frac{1}{\a'} \int_{b \a'}^{t_3 - c \a'} d t_2 +
\int_{0}^{c} d \tau_2~.
\label{arisplit}
\eeq
The $b$ and $c$ dependence arising from the upper limits of
integration of the $\tau$ integrals must cancel the dependence arising
from the $t_2$ integral when the limits of integration are expanded in
powers of $\a'$. The correct treatment of the $\tau_i$ integrations is
thus to perform the integral exactly, but retain only the contribution
from the lower limit of integration, $\tau_i = 0$, corresponding to
$y_i = 1$.  Notice that there is one (and only one) integration region
of this kind for every propagator in the loop, and the maximum number
of regions that can simultaneously contribute is correctly given by
the overall counting of powers of $\a'$ in the string master formula,
\eq{onemaster}. Notice also that the choice of integration variables
is built in the algorithm, since we are retaining only the contribution
of one limit of integration: a generic change of variables, in particular 
a shift by a constant, will in fact redistribute the result of the
integration between the two limits.

There is a class of terms contributing to scalar graphs with
four--point vertices that requires regularization, as was already
pointed out in Refs.~\cite{big} and \cite{maru}. These contributions
arise when one takes $z_{i + 1}$ close to $z_i$ in a term containing a
double derivative of the Green function precisely with respect to
these variables. Such terms diverge as $z_{i + 1} \to z_i$, {\it i.e.}
$y_i \to 1$, because of the double pole in $\partial_i \partial_{i +
1} G(z_i, z_{i + 1})$. To regularize this divergence, one may retain
also the $y$ dependence arising from the exponential of the same Green
function, which is formally suppressed in the $\a' \to 0$ limit. It's
easy to see that the resulting integral is of the form
\beq
I \left(c, \a' p_i \cdot p_{i + 1} \right) = \int_{e^{-c}}^{1} d y 
(1 - y)^{- 2 + 2 \a' p_i \cdot p_{i + 1}} y^{- \a' p_i \cdot p_{i + 1}}~,
\label{mafor}
\eeq
where $c$ is the arbitrary constant defining the boundary of the
appropriate integration region.  Notice that there is some freedom
concerning the inclusion of $y$--dependent terms arising from the
exponential in this regularization prescription: in \eq{mafor} we
included exactly only the $y$ dependence arising from $G(z_i, z_{i +
1})$, while all other nonsingular terms (arising from, say, $G(z_j,
z_{i + 1})$, with $j \neq i$) were dropped, to be treated according to
the general rules outlined at the beginning of this section. As we
will see in the next section, the inclusion of further $y$--dependent
terms in \eq{mafor} is equivalent to a reshuffling of
one--particle--reducible and irreducible contributions to the
amplitude.

The integral $I \left(c, \a' p_i \cdot p_{i + 1} \right)$ is linearly
divergent at the upper limit of integration when $\a' \to 0$, and
requires regularization. It is comforting to verify that three
independent regularization schemes give the same result in the field
theory limit. 
\begin{itemize}
\item The simplest approach is the one that was adopted in
Refs.~\cite{big} and~\cite{maru}. One can set $\a' \to 0$ {\it ab
initio} in \eq{mafor}, expand the integrand in powers of $y$, and
integrate the series term by term, obtaining
\beq
I (c, 0) = \int_{{\rm e}^{- c}}^{1} \frac{dy}{(1 - y)^2} =
\int_{{\rm e}^{- c}}^{1} \frac{dy}{y} \sum_{n = 1}^{\infty} n y^n =
\sum_{n = 1}^{\infty} y^n|^1_{{\rm e}^{- c}}~.
\label{mag1}
\eeq
The lower limit contribution depends on $c$, so it must be discarded,
as explained before. The remaining sum can be interpreted as a
$\zeta$--function, yielding
\beq
\sum_{n = 1}^{\infty} 1 = \lim_{s \rightarrow 0} \sum_{n = 1}^{\infty} 
n^{- s }= \zeta (0) = - \frac{1}{2}~.
\label{zreg}
\eeq
This $\zeta$--function regularization was used in \cite{maru} also at the 
two--loop level, giving the correct field theory limit.
\item Perhaps a more natural way to approach the regularization of
\eq{mafor} is to view it as an incomplete $B$ function. As we shall
see, this viewpoint nicely ties together diagrams with four--point
vertices with reducible diagrams. Since our answer must be
independent of $c$, we may turn the integral into an ordinary $B$
function by taking the $c \to \infty$ limit. Considering for
simplicity a contribution to the two--point amplitude, and thus
replacing $p_i \cdot p_{i + 1}$ with $-p^2$, we find
\beq
I \left(\infty, - \a' p^2\right) = B(- 1 - 2 \a' p^2, 
1 + \a' p^2)~.
\label{beta}
\eeq
After analytic continuation we get $I(\infty, 0) = - 1/2$, exactly as
in \eq{zreg}. 
\item Finally, at one loop, these results are confirmed by integration
by parts of double derivatives of the Green function, at the level
of the string master formula, as was done in \cite{big,beko}. After
integration by parts, the correct field theory limit is obtained by
considering only scalar graphs with cubic vertices, so that
effectively the singularity we are studying is removed. The
regularization given by \eq{zreg} turns out to be the correct
prescription to recover the same results without partial
integration.
\end{itemize}
Our working prescription will then be
\beq
I \left(c, - \a' p^2 \right) \to - \frac{1}{2}~.
\label{mafor2}
\eeq

\begin{figure}
\begin{center}
\begin{picture}(200,150)(0,-15)
\GlueArc(100,60)(30,-45,45){3}{3}
\GlueArc(100,60)(30,45,180){3}{4}
\GlueArc(100,60)(30,180,315){3}{4}
\LongArrowArcn(100,60)(20,180,45)
\Text(100,70)[]{$k$} 
\Gluon(50,80)(70,60){-3}{2} \Gluon(50,40)(70,60){3}{2}
\Gluon(122,81)(142,102){3}{2} \Gluon(122,38)(142,19){3}{2}
\Text(100,100)[]{$x_1$}
\Text(142,60)[]{$x_2 $}
\Text(100,20)[]{$x_3 $}
\Text(60,82)[b]{$p_2$}
\Text(142,82)[bl]{$p_3 $}
\Text(142,34)[l]{$p_4 $}
\Text(60,34)[]{$p_1 $}
\Text(45,85)[]{A}
\Oval(45,85)(7,7)(0)
\Text(45,35)[]{A}
\Oval(45,35)(7,7)(0)
\Text(147,107)[]{A}
\Oval(147,107)(7,7)(0)
\Text(147,14)[]{A}
\Oval(147,14)(7,7)(0)
\end{picture}
\end{center}
\caption{Four--point diagram with a quartic vertex: external gluons are 
background fields, marked by an A. Momenta $p_i$ are 
incoming.}\label{fourp12}
\end{figure}
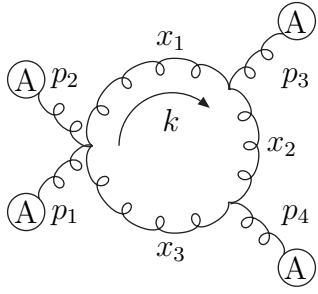

To illustrate the application of these rules, consider the diagram in 
\fig{fourp12}, with off--shell and non--transverse gluons.
%%%%%%%%%%%%%%%%%%%%%%%%%%%%%%%%%%%%%%%%%%%%%%%%%%%%%%%%%%%%%%%%%%%%%
%%%%%%%%%%%%%%%%%%%%%%%%%%%%%%%%%%%%%%%%%%%%%%%%%%%%%%%%%%%%%%%%%%%%%
To verify the correctness of our prescription, \eq{mafor2}, we can
focus on a specific kinematic structure, say $\e_1 \cdot \e_2~ \e_3
\cdot p_4 ~\e_4 \cdot p_2$\footnote{Notice that, in order to identify
univocally the coefficients, we write $\e_3 \cdot p_3 = - \e_3 \cdot
p_1 - \mbox{$\e_3 \cdot p_2$}- \e_3 \cdot p_4$ and $\e_4 \cdot p_4 =
-\e_4 \cdot p_1 - \e_4 \cdot p_2- \e_4 \cdot p_3$. The string master 
formula, in general, automatically generates the answer with this 
convention, which yields a particularly symmetric expression.}.  
Setting $z_2 = y$ in the master formula, \eq{onemaster} (recall that 
$z_1 = 1$), we obtain
\beqa
A^{(1)} (p_1, p_2, p_3, p_4) & \to & 8 N ~\Tr(\l^4 \l^3 \l^2 \l^1)
\frac{g_d^4}{(4 \pi )^{d/2}} (\e_1 \cdot \e_2)(\e_3 \cdot p_4)(\e_4
\cdot p_2) \nl
& \times & \int_0^{\infty } d t ~t^{- d/2} \int_0^t d t_4 
\int_0^{t_4} d t_3 \nl
& \times & \exp \left[ \frac{s_{1 2} t_3 (t - t_4)
-p_3^2 t_3 (t_4 - t_3) - p_4^2 (t_4 - t_3)(t - t_4)}{t}
\right] \nl
& \times & \frac{(2 t_4 - t)(t - 2 t_4 + 2 t_3)}{ 4 t^2} \label{ymagic} \\
& \times & \int^1 \frac{dy}{y} 
\left[ \left( \frac{1}{y} + y \right) + (d - 2) \frac{y}{(1 - y)^2} 
\right]~, \nonumber
\eeqa
where $s_{1 2} =- 2 p_1 \cdot p_2$, and we wrote only the relevant
limit of integration in $y$. The term proportional to $(1/y + y)$
integrates to zero, while for the term proportional to $y/(1 - y)^2$
we must use \eq{mafor2}. The result for the contribution of the chosen
kinematic structure to the scalar graph containing \fig{fourp12} is
\beqa
A^{(1)} (p_1, p_2, p_3, p_4) & \to &  N ~\Tr(\l^4 \l^3 \l^2 \l^1)
\frac{g_d^4}{(4 \pi )^{d/2}} (\e_1 \cdot \e_2)(\e_3 \cdot p_4)(\e_4
\cdot p_2) \nl
& \times & \int_0^{\infty } d t ~t^{- d/2+2} \int_0^1 d x_1 ~ d x_2~ 
d x_3~ \delta (1-\sum_{j=1}^3 x_j) \nl
 & \times  & \exp \left[ t\left( s_{1 2} x_1 x_3
- p_3^2 x_1 x_2 - p_4^2 x_2 x_3 \right)
\right] \nl
& \times & \left((2 - d)(1 - 2 x_2)(1 - 2 x_3) \right)~.
\label{ymag2}
\eeqa
The result is again in agreement with the computation in the
background field Feynman gauge, including the contribution of ghosts.

\sect{Reducible diagrams}
\label{Reg3}

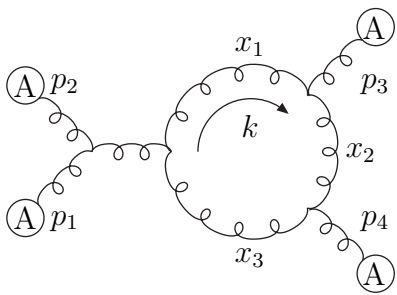
\begin{figure}
\begin{center}
\begin{picture}(160,150)(0,-15)
\GlueArc(100,60)(30,-45,45){3}{3}
\GlueArc(100,60)(30,45,180){3}{4}
\GlueArc(100,60)(30,180,315){3}{4}
\LongArrowArcn(100,60)(20,180,45)
\Text(100,70)[]{$k$} 
\Gluon(20,80)(40,60){3}{2} \Gluon(20,40)(40,60){3}{2}
\Gluon(122,81)(142,102){3}{2} \Gluon(122,38)(142,19){3}{2}
\Gluon(40,60)(70,60){3}{2}
\Text(100,100)[]{$x_1$}
\Text(142,60)[]{$x_2 $}
\Text(100,20)[]{$x_3 $}
\Text(30,82)[b]{$p_2$}
\Text(142,82)[bl]{$p_3 $}
\Text(142,34)[l]{$p_4 $}
\Text(30,34)[]{$p_1 $}
\Text(15,85)[]{A}
\Oval(15,85)(7,7)(0)
\Text(15,35)[]{A}
\Oval(15,35)(7,7)(0)
\Text(147,107)[]{A}
\Oval(147,107)(7,7)(0)
\Text(147,14)[]{A}
\Oval(147,14)(7,7)(0)
\end{picture}
\end{center}
\caption{Reducible four--point diagram: external gluons are background fields,
 marked by an A. Momenta $p_i$ are incoming.}\label{fourp12r}
\end{figure}

Reducible diagrams (such as the one portrayed in \fig{fourp12r}) arise
from the string master formula when two adjacent punctures, say $z_i$
and $z_{i + 1}$, are so close to each other on the world--sheet that
the corresponding correlator, $G(z_i, z_{i + 1})$, is dominated by its
logarithmic short distance singularity even before the field theory
limit is taken. In this integration corner, often called
`pinching' region \cite{beko}, the string master formula
effectively factorizes into the product of a one--loop amplitude with
an off--shell leg, times a tree--level amplitude. All string states may
flow through the propagator joining the two factors; the Yang--Mills
field theory limit is reached by selecting the poles corresponding to
the propagation of massless states. 

As we will see, reducible diagrams are the only diagrams where it is
necessary to require that a subset of the external legs (those forming
the tree that attaches to the irreducible loop) be on--shell. This
however is also a feature of diagrammatic computations in field theory
when using the background field method \cite{ags}: there, reducible
contributions to a scattering amplitude only coincide with the ones
obtained with conventional methods provided all external legs on at
least one side of the reducible propagator are put on shell.

The other relevant feature of the computation of reducible diagrams in
the background field method, namely the possibility of using different
gauges for the tree and the one--loop subamplitudes, is also preserved
in string theory. In fact, our analysis will confirm the results
obtained in \cite{bedu} for on--shell diagrams and in \cite{big} for
the divergent part of off--shell diagrams with transverse gluons: the
1PI part of the diagram is expressed in the background field Feynman
gauge, as expected from the previous sections; the tree part is
computed by the string in a different gauge, the Gervais--Neveu gauge.

The Gervais--Neveu gauge is a non--linear gauge, introduced to analyze
the field theory limit of open string theory \cite{gene}.  The
gauge--fixed action, ignoring ghosts, is:
\beq
S^{GN} = \int d^4 x \left( - \frac{1}{2} \Tr [F_{\mu\nu}^2] - \Tr 
[(\partial \cdot A - i g A^2)^2] \right)~.
\label{gn}
\eeq
The Feynman rules can be found, for example, in \cite{bernrev}. At
tree level, the vertices are simpler than those of the usual Feynman
gauge, allowing efficient calculations. At one loop, the string still
exploits this advantage in the tree part of the diagrams, but avoids
the complications that ghost terms would generate if the gauge were
used also for the one--loop part of the diagrams.

Reducible contributions can be extracted from the string master
formula by considering the region of moduli space where $z_i - z_{i +
m} \ll k \to 0$~\cite{beko,big}. To start with, we consider the case
$i = m = 1$, where, as in the previous section, $z_2 \to z_1 = 1$. In
this limit the string amplitude naturally factorizes in a sum over
poles which correspond to the propagation of different string states
in a `reducible' leg connecting two separate parts of the diagram.
We have already studied the contribution of the first term of this
sum, which is related to the exchange the tachyonic ground state.
With either of the two regularizations proposed, (\ref{zreg}) or
(\ref{beta}), one recovers contributions to  the irreducible diagram
depicted in Fig.~(\ref{fourp12}).

The basic idea of the `pinching' procedure adopted
in~\cite{beko,big} is to associate the contribution due to the
exchange of massless states to reducible diagrams.  In order to select
the appropriate pole for the massless exchange one has to perform a
Taylor expansions in $z_{i + j}$ around $z_i$ of all string Green
functions~(\ref{expgr}). However, for the `pinched' legs, which are
isolated on a reducible tree, we use an effective Green function
determined only by the short distance behavior of the string
expression, which is equal to the tree-level Green function,
\beq
G^{\rm E}(z_i,z_{j}) = \log |z_i - z_{j}|~.
\label{again}
\eeq
As an example, let us consider the reducible diagram in
\fig{fourp12r}, where  we have to assume that legs 1 and 2 are on
shell, as explained above. 
%%%%%%%%%%%%%%%%%%%%%%%%%%%%%%%%%%%%%%%%%%%%%%%%%%%%%%%%%%%%%%%%%%%%%
%%%%%%%%%%%%%%%%%%%%%%%%%%%%%%%%%%%%%%%%%%%%%%%%%%%%%%%%%%%%%%%%%%%%%
Taking the limit $z_2 \to 1$ by means of \eq{again}, we observe that 
the exponential involving $G(z_1, z_2)$ yields a factor of $(1 - 
z_2)^{- \alpha' s_{12}}$, while further factors of $(1 - z_2)$
can be generated by expanding the other Green functions and their
derivatives around $z_2 = 1$; for example,
\beqa
G_{32} & \stackrel{z_2 \to 1}{\longrightarrow} &  G_{31} + (1 - z_2)
\left[ - \frac{1 + z_3}{2 (1 - z_3)} + \frac{\log z_3}{\log k} + k
\frac{1 - z_3^2}{z_3} \right]\, , \nl
\partial _3 G_{32} & \stackrel{z_2 \to 1}{\longrightarrow} & \partial _3
G_{31} + (1 - z_2) \left[ - \frac{1}{(1 - z_3)^2} + \frac{1}{z_3 \log k} - k
\frac{(1 + z_3^2)}{z_3^2} \right]\, , \label{redexpand} \\
\partial _2 \partial _3 G_{32} & \stackrel{z_2 \to 1}{\longrightarrow} &
\partial _2 \partial _3 G_{32}|_{z_2 = 1} + (1 - z_2) 
\left[ - \frac{2}{(1 - z_3)^3} - \frac{1}{z_3 \log k} - 2 k
\right] \, , \nonumber 
\eeqa
where terms of order ${\cal O}((1 - z_2)^2)$ and ${\cal O}(k^2)$ have
been neglected.
Clearly, Taylor--expanding in powers of $(1 - z_2)$ generates a series
of poles in the $s_{12}$ channel; in fact, the general integral in
$z_2$ is of the form
\beq
\int_{z_3}^1 dz_2 (1 - z_2)^{- \a' s_{12} + a -1}\, ,
\label{z2int}
\eeq
with integer $a$. Since $z_3 \to 0$ exponentially as $\a'\to 0$, we can 
evaluate the integral (possibly by analytic continuation in the external 
momenta) as
\beq
\int_0^1 dz_2 (1 - z_2)^{- \a' s_{12} + a - 1} = (a - \a' s_{12})^{-1}~.
\label{pinchint}
\eeq
Different values of $a$ correspond to different mass eigenstates propagating
in the $s_{12}$ channel. For example, $a = -1$ corresponds to tachyon
propagation, which we now discard, while the expected gluon pole
$1/s_{12}$ is    
given by $a = 0$.  Notice that, as announced, we have to require $p_1$ and 
$p_2$ to be on shell in order to rewrite $2 p_1 \cdot p_2$ as $ - s_{12}$.   
Having isolated the massless pole, one may apply the procedure illustrated
in \secn{GeneralFT} and \secn{Reg1} (only cubic vertices are connected to 
the loop in this case) and get the correct expression for the diagram in 
\fig{fourp12r}. For example, selecting the term $(\e_1 \cdot \e_2)(\e_3 
\cdot p_4)(\e_4 \cdot p_2)$, one finds
\beqa
\lefteqn{\hspace{-1cm}A^{(1)}(p_1, p_2, p_3, p_4)  
\to \, g_d^4 N \Tr(\l^4 \l^3 \l^2 \l^1)
\frac{2}{(4 \pi )^{d/2}}
(\e_1 \cdot \e_2)(\e_3 \cdot p_4)(\e_4 \cdot p_2)} \\
& \times & \int_0^{\infty } \!dt \,\, t^{- d/2+2}
\int_0^1 \! dx_1~ dx_2~ d x_3~ \delta (1-\sum_{j=1}^3 x_j)\nl
& \times & \exp \left[ t (s_{12} x_1 x_3 -p_3^2 x_1 x_2-p_4^2 x_2 x_3)\right] 
\label{finred} \\
& \times & \frac{1}{s_{12}} \left\{ 2 (d-2) (1-2 x_2) t^{-1} -8[p_1\cdot
p_3+p_1\cdot p_4 (1-2 x_2)]~\right. \nl
& &~+~ \left. (d - 2) (1-2x_2)(1-2 x_3)[p_2\cdot p_3 (1-2 x_1)-p_2\cdot
p_4 (1-2 x_3)]  \right\}~. \nonumber
\eeqa

The approach just described has a simple diagrammatic interpretation
and, in particular, it has the advantage of clearly separating
reducible and irreducible contributions. However, the use of an {\it
ad--hoc} Green function, \eq{again}, may appear arbitrary and not
completely justified at the string level. Here we want to outline also
an alternative method for computing reducible diagrams, which is a
natural extension of our rules for the calculation of diagrams with
four--point vertices. As we will show, in this case one does not have
to introduce any effective Green functions. On the contrary, it is
sufficient to introduce the parameter $y = z_2/z_1$, as in \eq{regII},
and to carefully study the region $y = 1$. Using this second method, 
diagrams with quartic vertices and reducible diagrams are computed 
simultaneously.

The starting point of the alternative method is a natural
generalization of the regularization~(\ref{beta}) presented in
\secn{Reg2}. In particular, one has to retain all logarithmic
dependence on $y$ in the exponential containing the sum of Green
functions. In this way, one finds a term of the form
\beq
{\cal Y}(y) = (1 - y)^{2 \a' p_1 \cdot p_2} ~y^{- \a' p_1 \cdot p_2 +
\a' p_2 \cdot p_3 \left(1 - 2 \frac{t_3}{t} \right) + \a' p_2 \cdot p_4 
\left(1 - 2 \frac{t_4}{t} \right)}~,
\label{yy}
\eeq
which gives a natural regularization of the tachyonic divergence $(1 -
y)^2$. The $t$'s above are the Schwinger parameters related in the
usual way to the string variables: $t_3 = - \a' \log z_3$, $t_4 = -
\a' \log z_4$ and $t= - \a' \log k$.  As explained in \secn{Reg2}, for
regular terms we perform the $y$ integral exactly, keeping only the
contribution of the $y = 1$ limit of integration. In this case, the
factor~(\ref{yy}) has no effect on the integration. For singular
terms, we obtain a generalization of \eq{mafor}.  In particular, the
tachyonic double pole at $y \to 1$ gives the contribution
\beqa
& & \hspace{-3mm} \int_{e^{-c}}^1 dy~ (1 - y)^{-2} {\cal Y}(y) ~\to~
\int_{0}^1 dy~ (1 - y)^{-2} {\cal Y}(y) \nl 
& = & \hspace{-1mm} B \left(- 1 + 2 \a' p_1 \cdot p_2,
1 - \a' p_1 \cdot p_2 + \a' p_2 \cdot
p_3 \left(1 - 2 \frac{t_3}{t} \right) + \a' p_2 \cdot p_4 \left(1 - 2
\frac{t_4}{t} \right) \right) \nl
& = & - \frac{1}{2} - \frac{1}{2 p_1 \cdot p_2} \left[ p_2 \cdot
p_3 \left(1 - 2 \frac{t_3}{t} \right) + p_2 \cdot p_4 \left(1 - 2 
\frac{t_4}{t} \right) \right] + {\cal O}(\a')~.
\label{newbet}
\eeqa
Of course, \eq{newbet} is just a factor of the amplitude; to recover
the complete result one has to include the appropriate measure of 
integration and the exponential factor appearing in \eq{ymagic},
which encode the contribution coming from the punctures
not involved in the pinching procedure.  It is easy to see that the
first term reproduces the prescription we introduced in \eq{mafor2} to
get the irreducible diagram with a quartic vertex, \fig{fourp12}; the
other terms, containing a pole in $2 p_1 \cdot p_2$, contribute to the
reducible configuration, \fig{fourp12r}.

Simple poles in $(1 - y)$ must be treated in exactly the same way, 
obtaining again
\beq
\int_{e^{-c}}^1 dy ~(1 - y)^{-1} {\cal Y}(y) = \frac{1}{2 \a' p_1 \cdot p_2} 
+ {\cal O}((\a')^0)~,
\label{simplpl}
\eeq
which is another contribution to the reducible diagram.  Notice the
negative power of $\a'$ in \eq{simplpl}: it tells us that we need to
consider also contributions of order $\a'$, obtainable from the
expansion of the exponential, as outlined in \eq{expexp}; furthermore,
one must expand around $y = 1$ all the terms multiplying \eq{yy}, to
be sure to get all terms of the type~(\ref{simplpl}).  The difference
with the previous approach is that here we kept the exact dependence
on $y = z_2$ in all exponentiated Green functions and no expansion in
$y$ was done in the terms proportional to $\log(y)$, present, for
instance, in $G_{23}$.  Because of this, the contribution of the
simple pole is now different from the one of \eq{finred} and in
particular the last term proportional to $(d-2)$ is absent. The 
complete result is recovered only once one takes into account the 
reducible contribution coming from the new regularization~(\ref{newbet}).

The first method is maybe preferable from the computational point of
view, because it preserves the one--to--one correspondence between
string regions and Feynman graphs. The second method, on the other
hand, is interesting because it ties together the prescription for
quartic vertices, \eq{mafor2}, and the one for reducible diagrams.

It is not difficult to generalize the method we just presented to more
complicated situations where the pinching procedure involves two
non--consecutive legs, like $z_i$ and $z_{i + m}$. In this case, the
role of the variable $y$ is taken by the combination $z_{i + m}/z_i$.
Moreover, one can introduce the variables $\zeta_l =(z_{i + l} - z_{i
+ m})/(z_{i} - z_{i + m})$ for $l = 1,2, \ldots , m - 1$, which
factorize the expression of the amplitude in two separate parts. The
term depending only on the $\zeta$'s has the functional form of a
tree--level amplitude and describes the legs isolated in the reducible
part, while the other term of the amplitude can be treated as usual
and gives the contribution of the irreducible part.

\sect{Systematics}
\label{syst}

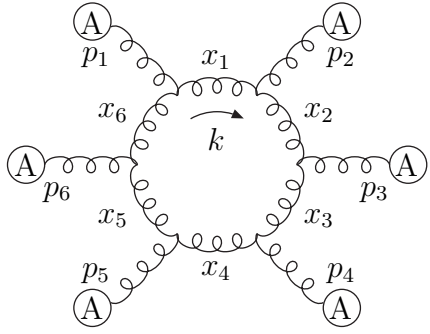
\begin{figure}
\begin{center}
\begin{picture}(200,150)(0,-15)
\GlueArc(100,60)(30,0,60){3}{3}
\GlueArc(100,60)(30,60,120){3}{3}
\GlueArc(100,60)(30,120,180){3}{3}
\GlueArc(100,60)(30,180,240){3}{3}
\GlueArc(100,60)(30,240,300){3}{3}
\GlueArc(100,60)(30,300,360){3}{3}
\LongArrowArcn(100,60)(20,120,60)
\Text(100,70)[]{$k$} 
\Gluon(60,112)(85,86){3}{3} \Gluon(115,86)(140,112){3}{3}
\Gluon(60,8)(85,34){-3}{3} \Gluon(115,34)(140,8){-3}{3}
\Gluon(35,60)(70,60){3}{3} \Gluon(130,60)(165,60){3}{3}
\Text(100,100)[]{$x_1$}
\Text(61,80)[]{$x_6 $} \Text(61,40)[]{$x_5 $}
\Text(139,80)[]{$x_2 $} \Text(139,40)[]{$x_3 $}
\Text(100,20)[]{$x_4 $}
\Text(60,97)[br]{$p_1$}
\Text(142,97)[bl]{$p_2 $} \Text(40,50)[]{$p_6 $}
\Text(142,19)[l]{$p_4 $}\Text(160,50)[]{$p_3 $}
\Text(60,19)[r]{$p_5 $}
\Text(53,114)[]{A}
\Oval(53,114)(7,7)(0)
\Text(53,6)[]{A}
\Oval(53,6)(7,7)(0)
\Text(147,114)[]{A}
\Oval(147,114)(7,7)(0)
\Text(147,6)[]{A}
\Oval(147,6)(7,7)(0)
\Text(28,60)[]{A}
\Oval(28,60)(7,7)(0)
\Text(172,60)[]{A}
\Oval(172,60)(7,7)(0)
\end{picture}
\end{center}
\caption{Six--point diagram: external gluons are background fields,
 marked by an A. Momenta $p_i$ are incoming.}\label{sixp}
\end{figure}

We have completed the list of all regions of string moduli space
contributing to the Yang--Mills field theory limit of the string
master formula. Each region corresponds to a scalar graph of a given
topology, and we have given a procedure to compute the corresponding
contribution to the field theory amplitude in each case. We emphasize
that all steps in the computation may be automated, in a manner
essentially independent of the number of external gluons. In the
special, but phenomenologically most relevant case of on--shell
scattering amplitudes, the steps of the calculation of a given
one--loop multigluon amplitude can be summarized as follows.
\begin{itemize}
\item Use color and helicity decomposition techniques to reduce the
computation of the full amplitude to its basic building blocks,
color--ordered fixed--helicity subamplitudes. For each independent
subamplitude, choose polarization vectors to minimize the number of
non--zero scalar products of polarizations and momenta.
\item List all scalar graphs contributing to the full amplitude, {\it
i.e} all one--loop graphs that would contribute to the corresponding
amplitude in a scalar theory with cubic and quartic vertices. Because
the answer will be given in dimensional regularization, graphs that
vanish in that scheme (such as tadpoles and bubbles on massless
external legs) need not be included.
\item Each subamplitude can now be written as a sum over scalar graphs,
\beqa
A^{(1)}_{\{\lambda_i\}, {\cal C}} ~(1, \ldots, M)  & = &
\frac{g_d^M}{(4 \pi)^{2 - \epsilon}} \int_0^\infty d t~ 
t^{- 2 + \epsilon} \sum_g \int_0^1 \prod_{i = 1}^{n_g} d x_i ~
\delta (1 - \sum_{j = 1}^{n_g} x_j)  \nl
& \times & \exp 
\Big[ - t S_g \left(x_i; p_i \cdot p_j \right) \Big] ~R_g 
\left(t, x_i; p_i, \epsilon_i^{(\lambda_i)} \right)~,
 \label{systamp}
\eeqa
where the subscript ${\cal C}$ denotes the color structure whose
coefficient we are computing, and $n_g$ is the number of propagators
in the scalar graphs $g$ contributing to the amplitude. The dynamical
information is encoded in the functions $S_g$ and $R_g$, which can be
computed according to the rules given in the previous sections.
Specifically, $S_g$ is the same function of external momenta and
Schwinger parameters that would appear in a scalar theory for a graph
with the same topology; thus, it can be easily computed in field
theory as well as derived from string theory. The non--trivial content
of Yang--Mills subamplitudes is given by the function $R_g$, which is
a sum of polynomials in the dimensionless Schwinger parameters $x_i$,
multiplied times integer powers of $t$. As a consequence, the $t$
integration can always be done at the outset, yielding a $\Gamma$
function. Using helicity techniques from the beginning will, in
general, greatly simplify the function $R_g$, which in fact will
vanish for many graphs for each subamplitude.
\item Perform Schwinger parameter integrals. At this stage, string
theory can offer no further help; techniques have been developed
\cite{beint}, however, that are sufficient to compute all one--loop
scalar integrals that might be needed for Yang--Mills amplitudes in
the foreseeable future.
\end{itemize}
The method just outlined allows, as is well known, to take advantage
of the best organization of the amplitude in terms of color and
helicity decompositions. The dimensional regularization scheme which 
naturally emerges in this approach is the so--called 't Hooft--Veltman 
scheme \cite{schemes}: polarizations of external, observed gluons are 
in four dimensions, while polarizations of gluons circulating in the 
loop or otherwise unobserved are in $d = 4 - 2 \epsilon$ dimensions, 
as indicated by the factor $(1 + k (d - 2))$ of \eq{3master}.
It should perhaps be further emphasized that a major practical
advantage for amplitudes with several gluons is the fact that loop
momentum integration has already been performed. To illustrate this
fact, consider the six--gluon diagram in \fig{sixp} (with all gluons
on--shell and transverse).
%%%%%%%%%%%%%%%%%%%%%%%%%%%%%%%%%%%%%%%%%%%%%%%%%%%%%%%%%%%%%%%%%%%%%
%%%%%%%%%%%%%%%%%%%%%%%%%%%%%%%%%%%%%%%%%%%%%%%%%%%%%%%%%%%%%%%%%%%%%

Using Feynman rules and introducing Schwinger parameters, this diagram
by itself generates approximately $10^4$ terms, each of which must
still be integrated over the loop momentum $k$.  Suppose we focus on a
particular kinematical structure, say $\e_1 \cdot \e_5 \e_2 \cdot \e_6
\e_3 \cdot p_5 \e_4 \cdot p_5$\footnote{Notice that for transverse
gluons not all products $\epsilon_i \cdot p_j$ for fixed $i$, $i
\neq j$, are independent. To pick a basis, we choose to eliminate the
products $\e_3 \cdot p_6$ and $ \e_4 \cdot p_6$ using momentum
conservation. The kinematic structure considered is present, for
example, in the six--gluon amplitudes with helicities fixed as $\e
^{-} (p_1 ,p_4), \e ^{-} (p_2 ,p_4), \e ^{-} (p_3 ,p_4), \e ^{+}
(p_4 ,p_3), \e ^{+} (p_5 ,p_3),\e ^{+} (p_6 ,p_3)$.}: for most of
the original $10^4$ terms, this kinematical structure is generated
only upon loop momentum integration, so while we can postpone
integration over Schwinger parameters, we must integrate all terms in
$k$. Methods to perform these integrals are well--known \cite{pv}, but
they generate larger and larger expressions as the power of $k$ grows;
in the present case, each contribution with $6$ powers of $k$ can
generate up to $6 \cdot 10^5$ terms. When all is said and done, the
coefficient of the chosen kinematical structure must be picked out of
more than $4 \cdot 10^6$ terms, and that applies to just one diagram.
Clearly, this is a time and memory consuming task.

The string technique avoids all this. The change of variables
identifying the field theory limit for a given scalar graph is known a
priori, and all subsequent computerized manipulations involve only
algebraic operations and series expansions. Using a symbolic
manipulation program, such as Mathematica, on a normal PC, the Feynman
parameter integrand for a given scalar graph, such as the one containing
\fig{sixp} and the corresponding ghost loops, can be obtained in
seconds. In this case, for example, computing the coefficient of the
chosen kinematic structure in the color--ordered amplitude multiplying
${\rm Tr} (\l^6 \ldots \l^1)$ yields
\beqa
S_g (x_i; p_i \cdot p_j) & = & 2 p_1 \cdot p_2 x_2 x_6 +
2 p_2 \cdot p_3 x_1 x_3 + 2 p_3 \cdot p_4 x_2 x_4 + 2 p_4 \cdot p_5 
x_3 x_5 \nl
& + & 2 p_5 \cdot p_6 x_4 x_6 + 2 p_6 \cdot p_1 x_1 x_5 + 
(p_1 + p_2 + p_3)^2 x_3 x_6 \nl
& + & (p_2 + p_3 + p_4)^2 x_1 x_4 +
(p_3 + p_4 + p_5)^2 x_2 x_5~,
\label{sg6}
\eeqa
and
\beqa
R_g (t, x_i; p_i, \epsilon_i ) & = & 16 t^3
x_5 \Big[ (d - 2)  x_5 - 2~t (p_1 \cdot p_3 + p_1 \cdot p_4 - p_2 
\cdot p_3 - p_2 \cdot p_4) \nl
& - & 4~t~x_5 (p_1 \cdot p_5 + p_2 \cdot p_6) \Big]
\e_1 \cdot \e_5 \e_2 \cdot \e_6 \e_3 \cdot p_5 \e_4 \cdot p_5 + \ldots~.
\label{rg6}
\eeqa
Needless to say, the result in \eq{rg6} agrees with the corresponding 
background field method calculation.

\sect{Concluding remarks}
\label{concl}

We have completed a systematic analysis detailing how the one--loop
string moduli space splits into particle graphs in the field theory
limit. For each class of graphs, a specific $\a'$--dependent change of
variables has been introduced, which parametrizes the limiting process
to be implemented on the integrand of the string master formula. We
have given a geometric prescription for the choice of projective
transformations associated with external particles, so that our
formulas apply to off--shell, arbitrarily polarized gluons, and can be
used directly to compute general gluon correlation functions, in a
manner which is completely alternative to, though much more efficient
than conventional Feynman rules, at least at one loop. Singularities
arising at the boundaries of different regions of moduli space have
been consistently regulated, so as to avoid double counting. The
method as it stands can be implemented in computer code using
available symbolic manipulation programs. Such a program would be able
to compute one--loop Yang--Mills amplitudes and gluon correlation
functions, starting from the string master formula, provided the
relevant Schwinger parameter integrals are known in dimensionally 
regularized form.

An obvious application of the method is the computation of one--loop
scattering amplitudes with several gluons, which are relevant for
next--to--leading order corrections to multijet production at hadron
colliders. Such amplitudes are known for up to five gluons \cite{bernfive};
the six--gluon amplitude remains quite challenging even with the present
technology, but it is a natural and interesting testing ground; in
particular, it could be used to construct finite predictions for NLO
partonic cross sections, since the seven--gluon tree--level amplitude
is known in closed form \cite{kk}.

Other possible applications arise from the fact the method can be
applied off shell. At tree level, amplitudes with one off--shell gluon
are related to currents \cite{bg}, which can be exploited to establish
recursion relations for the calculation of on--shell matrix
elements. For certain classes of diagrams, these recursion relation
have been generalized to one loop \cite{mah}, but the complete
extension including diagrams with gluon loops is not known. The
present formalism would appear to be a natural framework to pursue
this kind of generalization. Other areas of interest include the study
of the high--energy, collinear and infrared limits of the
amplitudes. For example, since one can now use string theory to
compute matrix elements uncontracted with polarization vectors, it
would be interesting to attempt a `stringy' derivation of the soft
gluon current, recently derived at one loop by conventional methods
\cite{cagra}.

Finally, we believe that the present systematic study is a necessary
step for the extension of the method to more than one loop. As was
shown in the case of scalar field theories, at two loops the
decomposition of string moduli space into particle graphs is much more
intricate than at one loop, even when considering only diagrams with
cubic vertices \cite{scal}. A thorough understanding of the one--loop
case on a graph by graph basis, including the non--trivial case of
diagrams with four point vertices, will be of considerable help in the
study of more complicated graph topologies.

\vskip 1cm

{\large {\bf {Acknowledgements}}}
\vskip 0.5cm

We would like to thank C. Schubert for several useful observations.
R.R. acknowledges the support  of the European Commission RTN programme
HPRN-CT-2000-00131 and of the Physics Department of the University of
Neuch\^atel, where part of this work was carried out.  L.M.
acknowledges the support of the European Commission TMR programme
FMRX-CT98-0194 (DG 12 - MIHT). Work supported in part by the italian
Ministero dell'Universit\`a e della Ricerca Scientifica.


\begin{thebibliography}{99}

\bibitem{nesche72} 
A.~Neveu and J.~Scherk, {\it Nucl.~Phys.} {\bf B36} (1972) 155.
%%CITATION = NUPHA,B36,155;%%

\bibitem{manpa} 
M.L.~Mangano and S.J.~Parke, {\it Phys.~Rep.} {\bf 200} (1991) 301.
%%CITATION = PRPLC,200,301;%%

\bibitem{beko}
Z.~Bern and D.A.~Kosower, {\it Phys.~Rev.}  {\bf D38} (1988) 1888;
{\it Phys.~Rev.~Lett.} {\bf 66} (1991) 1669; {\it Nucl.~Phys.} {\bf B379} 
(1992) 451;
%%CITATION = PHRVA,D38,1888;%%
%%CITATION = PRLTA,66,1669;%%
%%CITATION = NUPHA,B379,451;%%

\bibitem{bebo}
Z.~Bern, {\it Phys.~Lett.} {\bf B296} (1992) 85.
%%CITATION = PHLTA,B296,85;%%

\bibitem{bernrev}
Z.~Bern, in {\it Proceedings of TASI 92},
eds. J.~Harvey and J.~Polchinski, {\tt hep-ph/9304249};
%%CITATION = HEP-PH 9304249;%%
Z.~Bern, L.~Dixon and D.A.~Kosower, {\it Ann.~Rev.~Nucl.~Part.~Sci.} 
{\bf 46} (1996) 109, {\tt hep-ph/9602280}.
%%CITATION = HEP-PH 9304249;%%
%%CITATION = HEP-PH 9602280;%%

\bibitem{schu} For a review, see C.~Schubert ``Perturbative Quantum Field 
Theory in the String--Inspired Formalism'', preprint {\bf LAPTH-761/99}, 
to appear in {\it Phys.~Rep.}, and references therein.

\bibitem{nap}
L.~Cappiello, R.~Marotta, R.~Pettorino and F.~Pezzella, {\it Mod.~Phys.~Lett.}
{\bf A13} (1998) 2433, {\tt hep-th/9804032}; {\it Mod.~Phys.~Lett.}
{\bf A13} (1998) 2845, {\tt hep-th/9808164};
A.~Liccardo, R.~Marotta and F.~Pezzella, {\it Mod.~Phys.~Lett.}
{\bf A14} (1999) 799, {\tt hep-th/9903027};
F.~Cuomo, R.~Marotta, F.~Nicodemi, R.~Pettorino, F.~Pezzella and G.~Sabella,
{\tt hep-th/0011071}.
%%CITATION = HEP-TH 0011071;%%
%%CITATION = HEP-TH 9903027;%%
%%CITATION = HEP-TH 9808164;%%
%%CITATION = HEP-TH 9804032;%%

\bibitem{noncom} 
O.~Andreev and H.~Dorn, {\it Nucl.~Phys.}  {\bf B583} (2000) 145,
{\tt hep-th/0003113}; A.~Bilal, C.~Chu and R.~Russo,
{\it Nucl.~Phys.} {\bf B582} (2000) 65, {\tt hep-th/0003180}; C.~Chu, 
R.~Russo and S.~Sciuto, {\it Nucl.~Phys.} {\bf B585} (2000) 193,
{\tt hep- th/0004183}.
%%CITATION = HEP-TH 0003113;%%
%%CITATION = HEP-TH 0003180;%%
%%CITATION = HEP-TH 0004183;%%

\bibitem{bedu} 
Z.~Bern and D.C.~Dunbar, {\it Nucl.~Phys.} {\bf B379} (1992) 562.
%%CITATION = NUPHA,B379,562;%%

\bibitem{letter} 
P.~Di Vecchia, A.~Lerda, L.~Magnea and R.~Marotta, {\it Phys.~Lett.} 
{\bf B351} (1995) 445, {\tt hep-th/9502156}. 
%%CITATION = HEP-TH 9502156;%%

\bibitem{big} 
P.~Di Vecchia, L.~Magnea, A.~Lerda, R.~Russo and R.~Marotta,
{\it Nucl.~Phys.} {\bf B469} (1996)  235, {\tt hep-th/9601143}.
%%CITATION = HEP-TH 9601143;%%

\bibitem{2loop}
P.~Di Vecchia, L.~Magnea, A.~Lerda, R.~Marotta and R.~Russo,
{\it Phys.~Lett.} {\bf B388} (1996) 65, {\tt hep-th/9607141}.
%%CITATION = HEP-TH 9607141;%%

\bibitem{scal} 
A.~Frizzo, L.~Magnea and R.~Russo, {\it Nucl.~Phys.} {\bf B579} 
(2000) 379, {\tt hep- th/9912183}.
%%CITATION = HEP-TH 9912183;%%

\bibitem{mape}
R.~Marotta and F.~Pezzella, {\it Phys.~Rev.} {\bf D61} (2000) 106006, 
{\tt hep-th/9912158}; see also {\tt hep-th/0003044}.
%%CITATION = HEP-TH 9912158;%% 
%%CITATION = HEP-TH 0003044;%%

\bibitem{maru} L.~Magnea and R.~Russo, in {\it Proceedings} of ``DIS 97'',
Chicago, USA, 1997, eds. J.~Repond and D.~Krakauer, AIP Conf.~Proc. n.~407, 
913, {\tt hep-ph/9706396}; also in {\it Proceedings} of ``Beyond the 
Standard Model V'', Balholm, Norway, 1997, eds. G.~Eigen, P.~Osland and 
B.~Stugu, AIP Conf.~Proc. n.~415,   347, {\tt hep-ph/9708471}.
%%CITATION = HEP-PH 9706396;%%
%%CITATION = HEP-PH 9708471;%%

\bibitem{schmk} B.~Kors and M.G.~Schmidt, {\tt hep-th/0003171}. 
%%CITATION = HEP-TH 0003171;%%

\bibitem{Gomis:2000bn}
J.~Gomis, M.~Kleban, T.~Mehen, M.~Rangamani and S.~Shenker,
JHEP {\bf 0008} (2000) 011, {\tt hep-th/0003215}.
%%CITATION = HEP-TH 0003215;%%

\bibitem{copgroup} 
See, for example, P.~Di Vecchia, {\it ``Multiloop amplitudes in string
theory''} in Erice, {\it Theor.~Phys.} (1992), 16, and references therein.

\bibitem{GSW} 
M.B.~Green, J.H.~Schwarz and E.~Witten, 
{\it ``Superstring Theory''}, Cambridge University Press (1987).

\bibitem{copscho}
P.~Di Vecchia, F.~Pezzella, M.~Frau, K.~Hornfeck, A.~Lerda and
S.~Sciuto, {\it Nucl.~Phys.} {\bf B322} (1989) 317. 
%%CITATION = NUPHA,B322,317;%%

\bibitem{kaj} 
K.~Roland, {\it Phys.~Lett.} {\bf B289} (1992) 148.
%%CITATION = PHLTA,B289,148;%%

\bibitem{abbott} L.F.~Abbott, {\it Nucl.~Phys.} {\bf B185} (1981) 189.
%%CITATION = NUPHA,B185,189;%%

\bibitem{pv} G.~Passarino and M.~Veltman, {\it Nucl.~Phys.} {\bf B160}
(1979) 151.
%%CITATION = NUPHA,B160,151;%%

\bibitem{vanver} W.L.~van Neerven and J.A.M.~Vermaseren, {\it
Phys.~Lett.} {\bf B137} (1984) 241.
%%CITATION = PHLTA,B137,241;%%

\bibitem{bekoco} 
Z.~Bern and D.A.~Kosower, {\it Nucl.~Phys.} {\bf B362} (1991) 389.
%%CITATION = NUPHA,B362,389;%%

\bibitem{ags} L.F.~Abbott, M.T.~Grisaru and R.K.~Schaefer, {\it Nucl.~Phys.}
{\bf B229} (1983) 372.
%%CITATION = NUPHA,B229,372;%%

\bibitem{gene} J.L.~Gervais and A.~Neveu, {\it Nucl.~Phys.} {\bf B46} 
(1972) 381.
%%CITATION = NUPHA,B46,381;%%

\bibitem{beint} Z.~Bern, L.~Dixon and D.A.~Kosower, {\it Nucl.~Phys.} 
{\bf B412} (1994) 751, {\tt hep- ph/9306240}; T.~Binoth, J.Ph.~Guillet, 
G.~Heinrich, {\it Nucl. Phys.} {\bf B 572} (2000) 361, {\tt hep-ph/9911342}.
%%CITATION = HEP-PH 9306240;%%
%%CITATION = HEP-PH 9911342;%%

\bibitem{schemes} G.~'t~Hooft and M.~Veltman, 
{\it Nucl.~Phys.} {\bf B44} (1972) 189;
S.~Catani, M.H.~Seymour and Z.~Tr\'ocs\'anyi,  
{\it Phys.~Rev.} {\bf D55} (1997) 6819, {\tt hep-ph/9610553}. 

\bibitem{bernfive} Z.~Bern, L.~Dixon and D.A.~Kosower, {\it Phys.~Rev.~Lett.}
{\bf 70} (1993) 2677, {\tt hep-ph/9302280}.
%%CITATION = HEP-PH 9302280;%%

\bibitem{kk} R.~Kleiss and H.~Kuijf, {\it Nucl.~Phys.} {\bf B312} (1989) 616.
%%CITATION = NUPHA,B312,616;%%

\bibitem{bg} F.A.~Berends and W.~Giele, {\it Nucl.~Phys.} {\bf B306} (1988)
759.
%%CITATION = NUPHA,B306,759;%%

\bibitem{mah} G.~Mahlon, {\it Phys.~Rev.} {\bf D49} (1994) 4438, 
{\tt hep-ph/9312276}; in ``Beyond the Standard Model 4'', Lake Tahoe 1994,
475, {\tt hep-ph/9412350}.
%%CITATION = HEP-PH 9412350;%%

\bibitem{cagra} S.~Catani and M.~Grazzini, {\it Nucl.~Phys.} {\bf B591} 
(2000) 435, {\tt hep-ph/0007142}.
%%CITATION = HEP-PH 0007142;%%

 
\end{thebibliography}
\end{document}